\documentclass[conference]{IEEEtran}
\IEEEoverridecommandlockouts
\usepackage{cite}
\usepackage{amsmath,amssymb,amsfonts}
\usepackage{algorithmic}
\usepackage{graphicx}
\ifCLASSOPTIONcompsoc
\usepackage[caption=false,font=normalsize,labelfon
t=sf,textfont=sf]{subfig}
\else
\usepackage[caption=false,font=footnotesize]{subfi
g}
\fi
\usepackage{makecell}
\usepackage{textcomp}
\usepackage{xcolor}
\usepackage[misc]{ifsym}
\def\BibTeX{{\rm B\kern-. 05em{\sc i\kern-. 025em b}\kern-. 08em
    T\kern-. 1667em\lower. 7ex\hbox{E}\kern-. 125emX}}
    
\begin{document}

\title{Syndrome-aware Herb Recommendation with Multi-Graph Convolution Network
\thanks{\Letter Corresponding author. This work was supported by National Key R\&D Program of China (No. 2017YFC0803700), NSFC grants (No. 61532021 and No. 61972155), Shanghai Knowledge Service Platform Project (No. ZF1213), NSFC (No. 61702190, No. 61972372 and No. U19A2079) and Zhejiang Lab (No. 2019KB0AB04).}
}

 \author{\IEEEauthorblockN{Yuanyuan Jin\IEEEauthorrefmark{1}, Wei Zhang\IEEEauthorrefmark{1}, Xiangnan He\IEEEauthorrefmark{2}, 
Xinyu Wang\IEEEauthorrefmark{1} and \Letter Xiaoling Wang\IEEEauthorrefmark{1}\IEEEauthorrefmark{3}}
\IEEEauthorblockA{\IEEEauthorrefmark{1} Shanghai Key Laboratory of Trustworthy Computing, East China Normal University\\
 Shanghai, China \\
\{yyj, xinyuwang\}@stu.ecnu.edu.cn, zhangwei.thu2011@gmail.com, xlwang@cs.ecnu.edu.cn
 }
\IEEEauthorblockA{\IEEEauthorrefmark{2}School of Data Science,
University of Science and Technology of China\\
Hefei, China,
xiangnanhe@gmail.com}
\IEEEauthorblockA{\IEEEauthorrefmark{3}Shanghai Institute of Intelligent Science and Technology, Tongji University\\
 Shanghai, China}

 }


\maketitle
\begin{abstract}
 Herb recommendation plays a crucial role in the therapeutic process of Traditional Chinese Medicine  (TCM), which aims to recommend a set of herbs to treat the symptoms of a patient.  
 While several machine learning methods have
been developed for herb recommendation, they are limited in
modeling only the interactions between herbs and symptoms, and ignoring the intermediate process of syndrome induction. When performing TCM diagnostics, an experienced doctor typically induces syndromes from the patient's symptoms and then suggests herbs based on the induced syndromes. As such, we believe the induction of syndromes --- an overall description of the symptoms --- is important for herb recommendation and should be properly handled. 
However, due to the ambiguity and complexity of syndrome induction, most prescriptions lack the explicit ground truth of syndromes.
 
In this paper, we propose a new method that takes the implicit syndrome induction process into account for herb recommendation.
Specifically, given a set of symptoms to treat, we aim to generate an overall syndrome representation by effectively fusing the embeddings of all the symptoms in the set, so as to mimic how a doctor induces the syndromes.
Towards symptom embedding learning, we additionally construct a symptom-symptom graph from the input prescriptions for capturing the relations  (co-occurred patterns)  between symptoms; we then build graph convolution networks  (GCNs)  on both symptom-symptom and symptom-herb graphs to learn symptom embedding.  
Similarly, we construct a herb-herb graph and build GCNs on both herb-herb and symptom-herb graphs to learn herb embedding, which is finally interacted with the syndrome representation to predict the scores of herbs. 
The advantage of such a Multi-Graph GCN architecture is that more comprehensive representations can be obtained for symptoms and herbs.   
We conduct extensive experiments on a public TCM dataset, demonstrating significant improvements over state-of-the-art herb recommendation methods.  Further studies justify the effectiveness of our design of syndrome representation and multiple graphs.

\end{abstract}


\section{Introduction}
\label{intro}

\begin{figure}[!tb]
  \centering
  \includegraphics[width=0.9\linewidth]{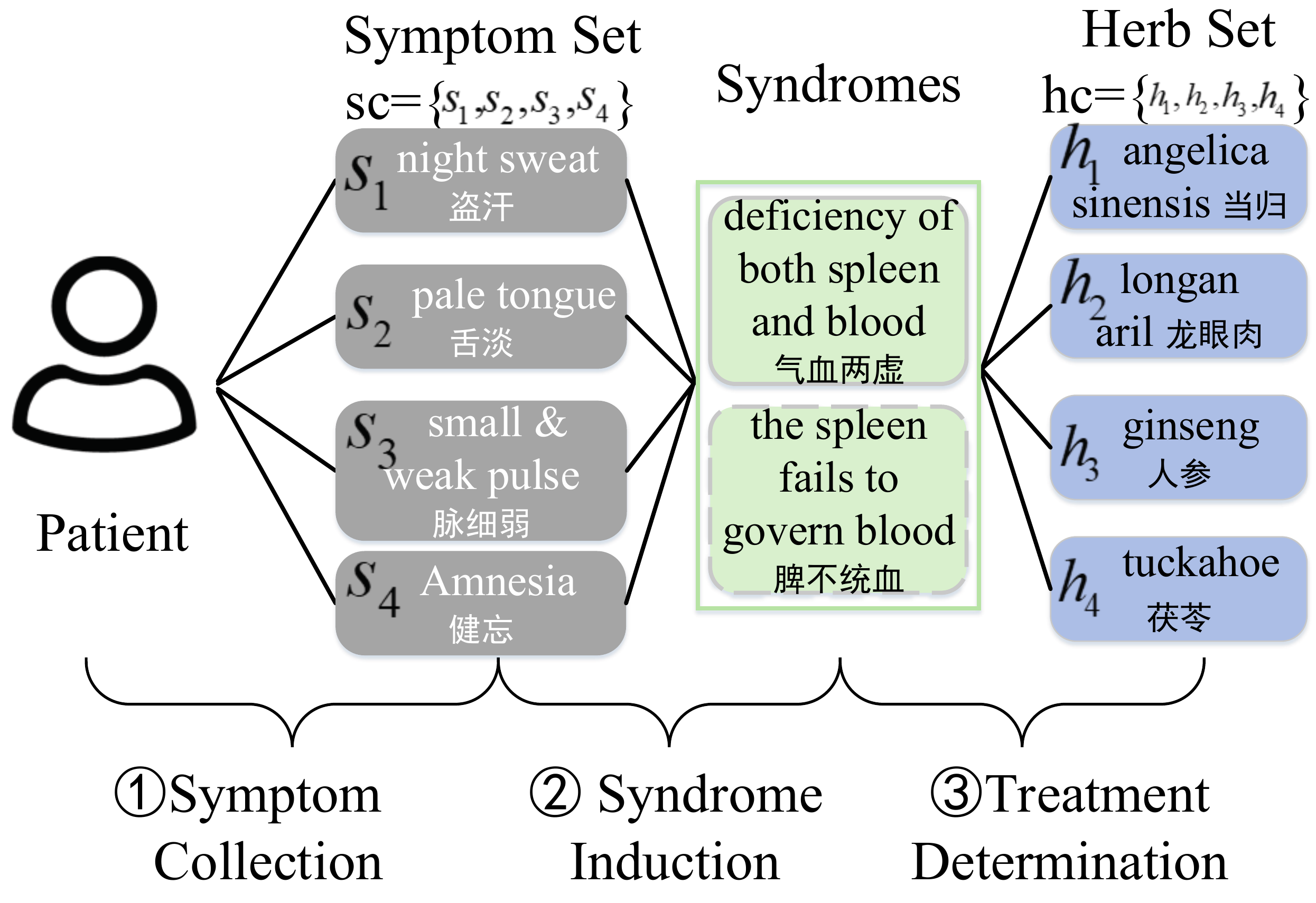} 
  \caption{An example of the therapeutic process in TCM. }
  \vspace{-5pt}
  \label{fig:treat} 
\end{figure}

As an ancient and holistic treatment system established over thousands of years, Traditional Chinese Medicine  (TCM)  plays an essential role in Chinese society \cite{cheung2011tcm}. 
The basis of the TCM theory is the thinking of holism, which emphasizes the integrity of the human body and its interrelationship with natural environments \cite{Wang2014Zheng}. 
Fig.~\ref{fig:treat} takes the classic \emph{Guipi Decoction} prescription as an example to show the three-step therapeutic process in TCM:
 (1)  \emph{Symptom Collection}.  The doctor examines the symptoms of the patient. Here the symptom set $sc$ contains ``night sweat'', ``pale tongue'', ``small and weak pulse'' and ``amnesia''.  
 (2)  \emph{Syndrome Induction}.  Corresponding syndromes are determined after an overall analysis of symptoms. In this case, the main syndrome is `` deficiency of both spleen and blood'' in a solid circle. As ``pale tongue'' and ``small and weak pulse'' can also appear under  ``the spleen fails to govern blood'', there is also an optional syndrome called ``the spleen fails to govern blood'' in a dotted circle. 
 (3)  \emph{Treatment Determination}.  The doctor chooses a set of herbs as the medicine to cure the syndromes. The compatibility of herbs is also considered in this step. 
Here the herb set $hc$ consists of ``ginseng'', ``longan aril'',  ``angelica sinensis'' and  ``tuckahoe''.  
As we can see, the second step of syndrome induction, which systematically summarizes the symptoms, is very critical for the final recommendation of herbs. However, as the above example shows, 
a symptom can appear in various syndromes, which makes the syndrome induction ambiguous and complex \cite{TCM}. Actually, for a certain symptom set, different TCM doctors might give different syndrome sets (as shown in Fig.~\ref{fig:treat}), and thus no standard ground truth exists.

In a TCM prescription corpus, each data instance contains two parts --- a set of symptoms and a set of herbs, which means the herb set can well cure the symptom set.  
To generalize to the unseen symptom set, the herb recommendation task focuses on modeling the interactions between symptoms and herbs, which is analogous to the traditional recommendation task that models the interactions between users and items~ \cite{wang2019neural}.  
Notably, one key difference is that in traditional recommendation, the prediction is mostly performed on the level of a single user, whereas in the herb recommendation, we need to jointly consider a set of symptoms to make prediction. 
Due to the lack of ground truth syndromes,  existing efforts on herb recommendation \cite{Yao2018ATM,ji2017latent,Ruan2019DiscoveringRF,ruan2019exploring} treat the syndrome concept as latent topics. However, they only learn the latent syndrome topic distribution given a single symptom.
Particularly, they focus on modeling the interaction between one symptom and one herb; and then the interactions from multiple symptoms are aggregated to rank the herbs. 
As such, the set information of symptoms is overlooked.

In this paper, we propose to incorporate the implicit syndrome induction process into herb recommendation, which conforms to the intuition that syndrome induction is a key step to summarize the symptoms towards making effective herb recommendations, as shown in Fig.~\ref{fig:treat}.
Specifically, given a  set of symptoms to treat,  we aim to learn an overall implicit syndrome representation based on the constituent symptoms, before interacting with herbs for recommendation generation.
Through this manner, the prescription behavior of doctors could be mimicked.

To this end, we propose a new method named \textit{Syndrome-aware Multi-Graph Convolution Network}  (SMGCN), a multi-layer neural network model that performs interaction modeling between syndromes and herbs for the herb recommendation task.  
In the interaction modeling component  (top layer of SMGCN), we first fuse the embeddings of the symptoms in a target symptom set via a Multi-Layer Perceptron (MLP) to directly obtain the overall implicit syndrome representation, which is later interacted with herb embeddings to output prediction scores.
In the embedding learning component  (bottom layers of SMGCN), we learn symptom embedding and herb embedding via GCN on multiple graphs.  Specifically, in addition to the input symptom-herb graph, we further build symptom-symptom and herb-herb graphs based on the co-occurrence of symptoms  (herbs)  in prescription entries.  
Intuitively, some symptoms are frequently co-occurred in patients  (e.g., nausea and vomit), modeling which is beneficial to symptom representation learning; similarly, the herb-herb graph evidences the frequently co-occurred herbs, which are useful for encoding their compatibility.  
We conduct experiments on a public TCM dataset~\cite{Yao2018ATM}, demonstrating the effectiveness of our SMGCN method as a whole and validating the rationality of each single purposeful design.

The main contributions of this work are as follows.  

\begin{itemize}
    \item We highlight the importance of representing syndromes and modeling the interactions between syndromes and herbs for herb recommendation.  
    \item We propose SMGCN, which unifies the strengths of MLP in fusion modeling  (i.e., fusing symptom embeddings into the overall implicit syndrome embedding)  and GCN in relational data learning (i.e., learning symptom and herb embeddings) for herb recommendation.  
    \item We build herb-herb and symptom-symptom graphs to enrich the relations of herbs and symptoms, and extend GCN to multiple graphs to improve their representation learning quality.

\end{itemize}

The rest of the paper is organized as follows: Section 2 describes the problem definition.  Section 3 describes our overall framework.   Section 4 introduces our proposed method.  Section 5 evaluates our method.  Section 6 surveys the related work.   Finally, section 7 provides some concluding remarks.

\section{PROBLEM DEFINITION}\label{AA}
The task of herb recommendation aims to generate a herb set as the treatment to a specific symptom set.
Herb recommender systems usually learn from the large prescription corpus.  Let $S$=$\{s_{1}, s_{2}, . . . ,s_{M}\}$ and $H$=$\{h_{1}, h_{2}, . . . , h_{N}\}$ denote all symptoms  and  herbs, respectively.  Each prescription consists of a symptom set and a herb set, e.g., $p$=$\langle\{s_{1}, s_{2}, . . .\}, \{ h_{1}, h_{2}, . . . \}\rangle$. 
In the syndrome induction process, an overall syndrome presentation needs to be induced for each symptom set, which is later used to generate an appropriate herb set. 
Hereafter we represent \textbf{symptom set} and  \textbf{herb set} by  \textbf{$sc$=$\{ s_{1}, s_{2}, . . . \}$} 
and \textbf{$hc$=$\{ h_{1}, h_{2}, . . . \}$}, respectively. 
In this way, each prescription is denoted by  $p$=$\langle sc, hc\rangle$.

Given a symptom set $sc$, our task is to compute an N-dimensional probability vector, where the value of dimension $i$ represents the probability that herb $i$ can cure $sc$. 
This is achieved by a learned prediction function $\hat{y}_{sc}$=$g (sc,H; \theta) $, where $\hat{y}_{sc}$ represents  the probability vector, and $\theta$ indicates the trainable parameters of function $g$. 
The input and output are defined as follows:
\begin{itemize}
    \item Input: Herbs $H$, Symptoms $S$,   Prescriptions $P$. 
    
    \item Output: A learned function $g (sc, H; \theta) $, which generates the probability vector $\hat{y}_{sc}$ for all herbs from $H$ given the symptom set $sc$. 
\end{itemize}

\begin{figure*}[htbp]
  \centering
  \includegraphics[width=1.0\linewidth]{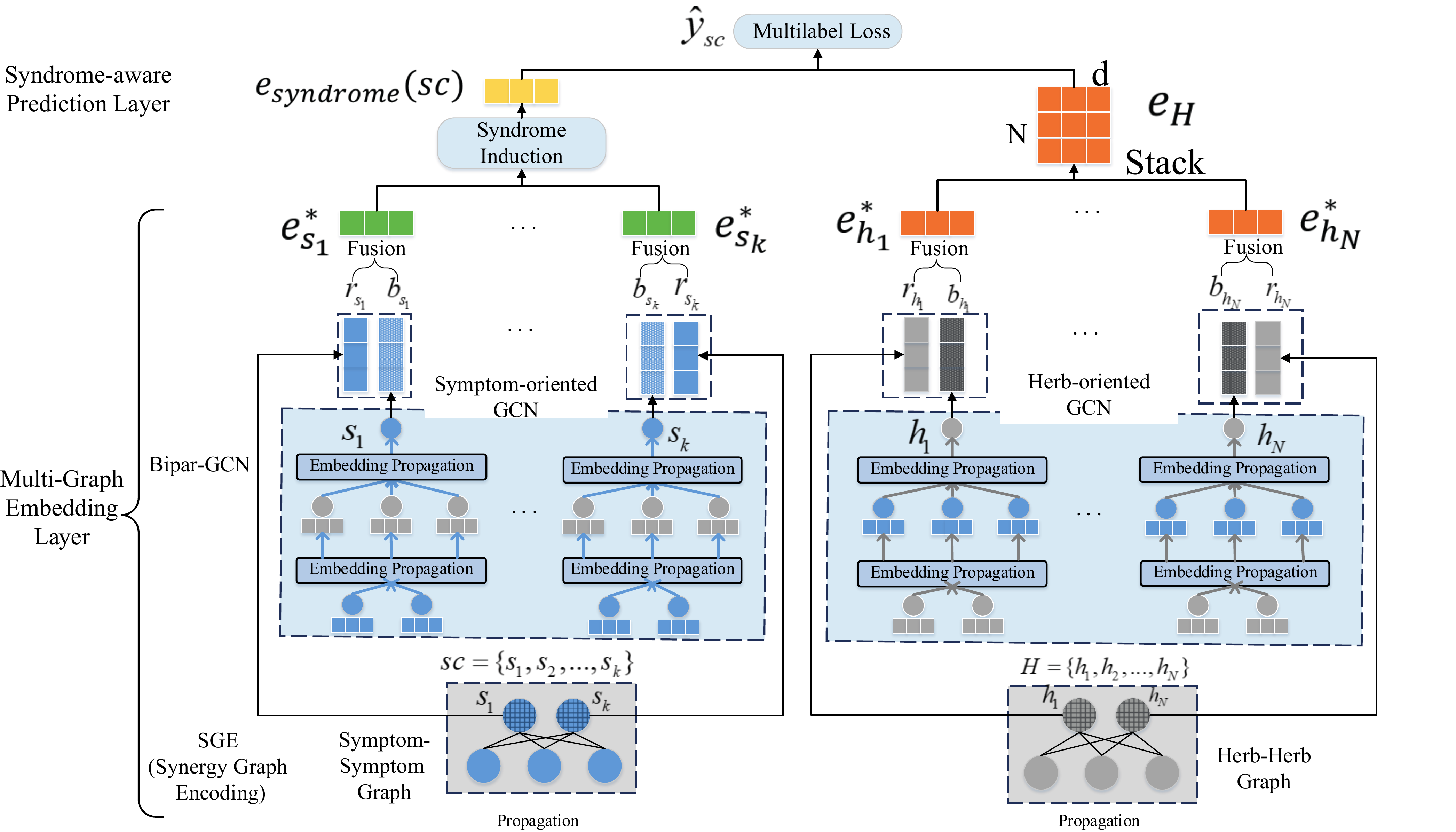}
  \caption{The overall architecture of our proposed model  (including Bipar-GCN, Synergy Graph Encoding  (SGE)  and Syndrome Induction  (SI) ).  Symptom nodes are in blue and herb nodes are in gray. Notably, the nodes with oblique lines are target nodes.} 
  \vspace{-5pt}
  \label{fig:framework}
\end{figure*}

\section{OVERVIEW OF PROPOSED APPROACH}

In this section, we discuss the proposed \textbf{Syndrome-aware Multi-Graph Convolution Network}  framework in detail, which is depicted in Fig. \ref{fig:framework}.  Our proposed model takes a symptom set $sc=\{s_{1}, s_{2},. . . , s_{k}\}$ and all herbs $H=\{h_{1},. . . , h_{N}\}$ as input, and outputs the predicted probability vector $\hat{y}_{sc}$ in dimension  $|H|$.  In $\hat{y}_{sc}$, the value at position $i$ indicates the probability that $h_{i}$ is appropriate to cure  $sc$.  

To complete this task, it mainly consists of two layers: the Multi-Graph Embedding Layer and Syndrome-aware Prediction Layer.

 \textbf{Multi-Graph Embedding Layer. } This layer aims to obtain expressive representations for all symptoms from $S$ and all herbs from $H$.  Considering the complex interrelations between symptoms and herbs in TCM, we first develop a Bipartite Graph Convolutional Neural Network  (Bipar-GCN) to process the bipartite symptom-herb graph.  To capture the intrinsic difference between symptoms and herbs, Bipar-GCN performs symptom-oriented embedding propagation for the target symptom node, and herb-oriented embedding propagation for the target herb node, respectively.
 Through this way, symptom embedding $b_{s}$ and herb embedding $b_{h}$ are learned. Second, a Synergy 
 Graph  Encoding (SGE)  component is introduced to encode the synergy information of symptom pairs and herb pairs.   For symptom pairs, it constructs a symptom-symptom graph according to the concurrent frequency of symptom pairs and performs the graph convolution on the symptom-symptom graph to learn symptom embedding $r_{s}$.  Analogously, it also learns herb embedding $r_{h}$ from a herb-herb graph. Third, for each symptom (herb),  two types of embeddings $b$  and $r$ from the Bipar-GCN and SGE  are fused to form integrated embeddings $e^{*}$. 
     
\textbf{Syndrome-aware Prediction Layer. } In this layer, bearing the importance of syndrome induction process in mind, the Syndrome Induction (SI)  component feeds the embeddings of all symptoms in the symptom set $sc$ into an MLP to generate the overall syndrome representation $e_{syndrome}(sc)$.  Second, all herb embeddings are stacked into $e_{H}$, an $N \times d$ matrix where $d$ is the dimension of each herb embedding.   The syndrome representation $e_{syndrome}(sc)$  interacts  with $e_{H}$ to predict $\hat{y}_{sc}$, the probability score vector for all herbs from $H$.

Considering that a set of herbs will be recommended as a whole, a multi-label loss function is utilized to optimize our proposed model.  All notations used in this paper are summarized in Tab.~\ref{tab:notaion_list}.  

\setcounter{table}{0}

\begin{table}[!tb]
\caption{Summary of all notations }
\centering
\begin{tabular}{|l|l|}
\hline
$e_{h}, e_{s}$ & initial embeddings for herbs, symptoms\\ \hline
$S, H$         & symptom collection and herb collection\\ \hline
$SC, HC$     & \begin{tabular}[c]{@{}l@{}}collection of symptom sets \\ and collection of herb sets\end{tabular}\\ \hline
$P$            & prescription collection\\ \hline
$N_{s}, N_{h}$ & \begin{tabular}[c]{@{}l@{}}neighborhood of symptom, herb \\ on the bipartite graph\end{tabular}\\\hline
$SH$ & symptom-herb graph \\\hline
$ SS, HH $      & \begin{tabular}[c]{@{}l@{}} symptom-symptom graph and\\ herb-herb graph\end{tabular} \\ \hline
$x_{s}, x_{h}$ &  threshold for constructing SS and HH \\ \hline
$N_{s}^{SS}, N_{h}^{HH}$  & \begin{tabular}[c]{@{}l@{}}neighborhood of symptom on SS   \\ neighborhood of herb on HH\end{tabular}    \\ \hline
$T_{s}^{k}, T_{h}^{k}$          & \begin{tabular}[c]{@{}l@{}}message construction function for symptom, herb  \\ at k-th Bipar-GCN layer\end{tabular}          \\ \hline
$W_{s}^{k}, W_{h}^{k}$           & \begin{tabular}[c]{@{}l@{}}message aggregation function for symptom, herb  \\ at k-th Bipar-GCN layer\end{tabular}           \\ \hline
$V_{s} , V_{h}$                                                & \begin{tabular}[c]{@{}l@{}}aggregation function for symptom on SS    \\  aggregation function for herb on HH\end{tabular} \\ \hline
$b_{N_{s}}^{k}, b_{N_{h}}^{k}$ & \begin{tabular}[c]{@{}l@{}}symptom, herb neighborhood embedding  \\ at k-th Bipar-GCN layer\end{tabular}                     \\ \hline
$b_{s}^{k}, b_{h}^{k}$           & \begin{tabular}[c]{@{}l@{}}symptom, herb output embeddings at \\ k-th Bipar-GCN   layer  \end{tabular}                                                                      \\ \hline
$r_{s}, r_{h}$                                                 & \begin{tabular}[c]{@{}l@{}}symptom  output embeddings on SS  \\ herb  output embeddings on HH\end{tabular}  \\ \hline
$e_{h}^{*}, e_{s}^{*}$  & herb, symptom final embedding after fusion \\ \hline
$W^{mlp}, b^{mlp}$ & \begin{tabular}[c]{@{}l@{}}the MLP weight matrix and bias parameter  \\ used in Syndrome Induction\end{tabular}  \\ \hline
$W^{att}, z$  & the attention network parameters in HeteGCN \\ \hline
 
$e_{syndrome}(sc)   $                         & \begin{tabular}[c]{@{}l@{}}the induced syndrome embedding \\ for symptom set  sc              \end{tabular}                 \\ \hline
$\hat{y}(sc)$  & \begin{tabular}[c]{@{}l@{}}the predicted probability  vector \\for $sc$  in  dimension $|H|$\end{tabular}    \\ \hline

\end{tabular}
    \label{tab:notaion_list}
\end{table}

\section{METHODOLOGIES}
 
\subsection{Bipartite Graph Convolution Network} 
\begin{figure}[!tb]
    \centering
    \includegraphics[width=0.8\linewidth]{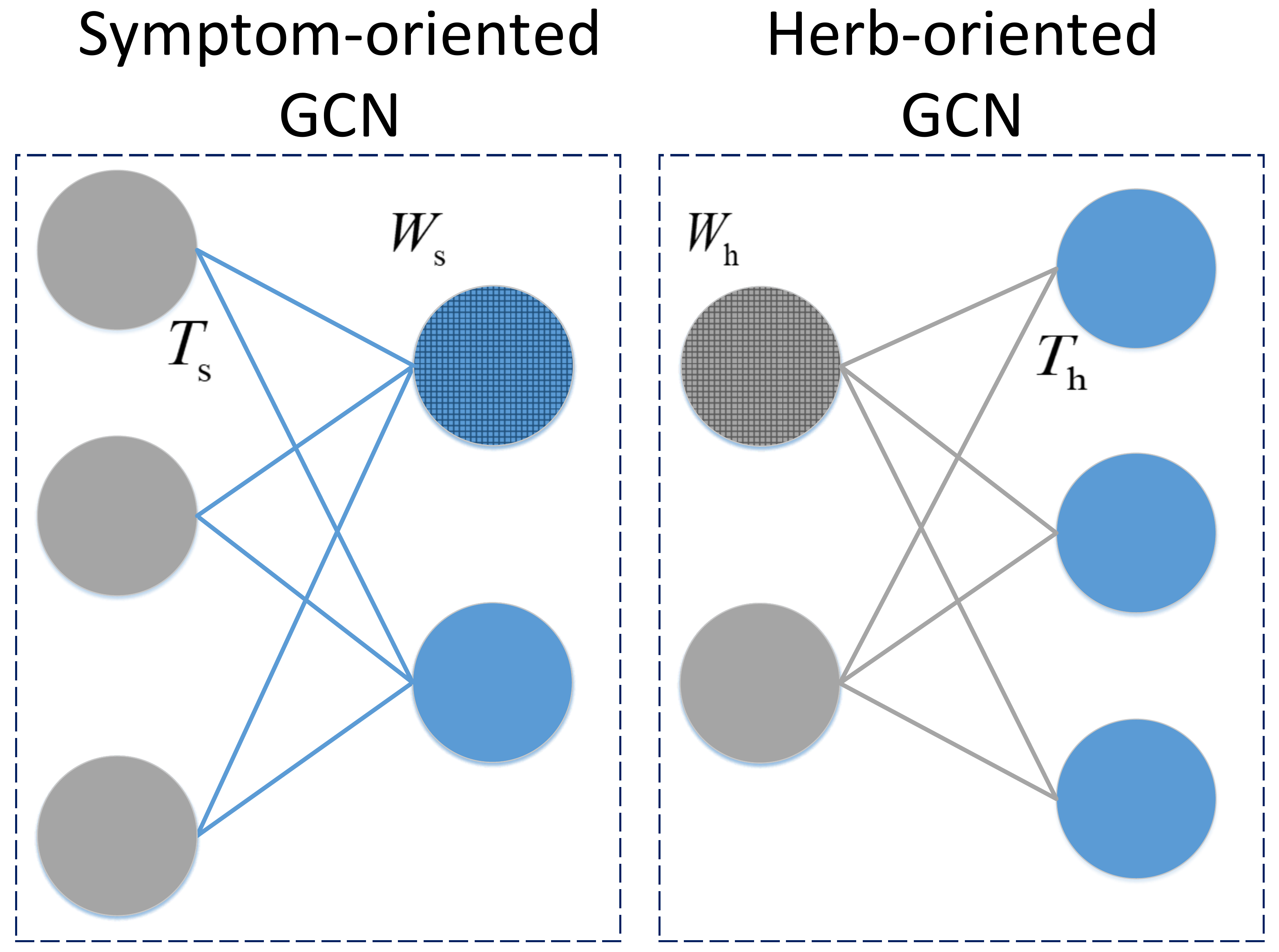}
    \caption{Bipartite GCN.  Blue edges and gray edges denote different graph convolution functions. The nodes with oblique lines are target nodes. }
    \label{fig:bipar-gcn}
\end{figure}

Recent works like  \cite{wang2019neural} have demonstrated the convincing performance of performing graph convolutions on the user-item graph in recommender systems. 
Despite their effectiveness, we argue that they ignore the intrinsic difference between the two types of nodes (users and items)  in the bipartite graph and employ a shared aggregation and transformation function across the graph, which may restrict the flexibility of information propagation and affect the embedding expressiveness to some extent. 
To model the intrinsic difference between herbs and symptoms, we leverage Bipar-GCN, which is shown in Fig.~\ref{fig:bipar-gcn}.  When the type of the target node is  ``symptom'', the left \emph{Symptom-oriented GCN} will be used to obtain the representation for this target node. Otherwise,  the right  \emph{Herb-oriented GCN} is adopted to learn the node embedding.  These two parts share the same topological structure of the symptom-herb graph but adopt different aggregation and transformation functions.   
In this way, different types of nodes can develop their own propagation flexibility and therefore learn more expressive representations.  Next we will introduce Bipar-GCN in detail. 
\subsubsection{Symptom-Herb Graph Construction}
Taking a TCM prescription $p$=$\langle sc$=$\{ s_{1}, s_{2}, . . . ,s_{k}\}, hc$=$\{ h_{1}, h_{2}, . . . , h_{m}\}\rangle$ as an example, symptoms and herbs in the same prescription are related to  each other. Therefore, $\{ (s_{1},h_{1}),. . . , (s_{1},h_{m}),. . . ,  (s_{k},h_{1}),. . . , (s_{k},h_{m}) \}$ constitute graph edges.
We take the symptom-herb graph as an undirected graph which is formulated as follows:
$$ SH_{s,h}, SH_{h,s}=\left\{
\begin{aligned}
1, &  ~\text{if } (s, h) \text{ co-occur in prescriptions}; \\
0, &  ~\text{otherwise},
\end{aligned}
\right. 
$$
wherein $SH$ indicates the symptom-herb graph.

\subsubsection{Message Construction}
In order to  propagate information from each neighbor node to the target node, there are two operations to be defined: how to generate information that each node transfers to the target node and how to merge multiple neighbor messages together. 

For symptom $s$, the message its one-hop neighbor herb $h$ transfers to it is defined as $m_{h}$,
\begin{equation} 
m_{h}^{0} =e_{h} \cdot T_{s}^{1}, 
\label{info}
\end{equation}
where $e_{h}$ is the initial embedding of herb $h$.  $T_{s}^{1}$ is the transformation weight matrix of the first-layer  (symptom). 
After collecting messages from all neighbors, we choose average operation to merge them, which is defined as follows,
\begin{equation} 
\label{sym_equation}
b_{N_{s}}^{0} = tanh (\frac{1}{|N_{s}|}\sum_{h\in N_{s}}m_{h}^{0}), 
\end{equation}
where $N_{s}$ is the one-hop neighbor set of $s$ and we choose $tanh$ as the activation function. 
Analogously, for herb $h$, the merged one-hop neighbor message can be represented by,
\begin{equation}\label{eq:ave-agg}
b_{N_{h}}^{0} = tanh (\frac{1}{|N_{h}|}\sum_{s\in N_{h}}m_{s}^{0}), 
\end{equation}
where $m_{s}^{0} =e_{s} \cdot T_{h}^{1}$. 

\subsubsection{Message Aggregation}
After receiving the merged neighbor representation, the next step is to update the embedding for the target node.  Here we adopt the $GraphSAGE$ $Aggregator$ proposed in  \cite{Hamilton2017InductiveRL}, which concatenates two representations, followed by a nonlinear activation operation.  The first-layer  symptom representation $b_{s}^{1}$ and herb representation $b_{h}^{1}$ are defined as follows,
\begin{equation} 
\label{b_s_1}
b_{s}^{1} = tanh (W_{s}^{1}\cdot  (e_{s} || b_{N_{s}}^{0}) ), 
\end{equation}
\begin{equation} 
b_{h}^{1} = tanh (W_{h}^{1}\cdot  (e_{h} || b_{N_{h}}^{0}) ), 
\end{equation}
where $||$ indicates the concatenation operation of two vectors.  $W_{s}$ and $W_{h}$ denote the aggregation weight matrices for symptoms and herbs, respectively. 

\subsubsection{High-order Propagation}
We can further extend the one-hop propagation rule to multiple layers. 
Specifically, in the k-th layer, we recursively formulate the representation of herb $h$ as,
\begin{equation} 
b_{h}^{k} = tanh (W_{h}^{k}\cdot  (b_{h}^{k-1} || b_{N_{h}}^{k-1}) ), 
\end{equation}
wherein the message from neighbors in the k-th layer   for  $h$ is defined as follows,
\begin{equation} 
b_{N_{h}}^{k-1} =tanh (\frac{1}{|N_{h}|}\sum_{s\in N_{h}}b_{s}^{k-1} \cdot T_{h}^{k}),
\end{equation}
For symptom $s$, the formulations are similar,
\begin{equation} 
b_{s}^{k} = tanh (W_{s}^{k}\cdot  (b_{s}^{k-1} || b_{N_{s}}^{k-1}) ),
\label{symprop}
\end{equation}
wherein the message propagated within k-th layer for  $s$ is defined as follows,
\begin{equation} 
b_{N_{s}}^{k-1} =tanh (\frac{1}{|N_{s}|}\sum_{h\in N_{s}}b_{h}^{k-1} \cdot T_{s}^{k}). 
\end{equation}

\subsection{Synergy Graph Encoding Layer}

Except for the symptom-herb relation, there are also some synergy patterns within symptoms and herbs.  Given a prescription $p$=$\langle sc, hc\rangle$, symptoms in $sc$ are not independent but related to each other, and herbs in $hc$ also influence each other and form a complete composition. 
As such, these relations could be exploited to construct synergy graphs for symptoms and herbs, respectively.
It is worth noting that although the two-order information propagation on the symptom-herb graph can capture the homogeneous relations between herbs and symptoms, the second-order symptom-symptom and herb-herb links are not equal to the concurrent pairs in prescriptions. For example, in prescriptions $p_{1}$=$\langle \{s_{1}, s_{2}\}, \{h_{1}, h_{2}\}\rangle$ and $p_{2}$=$\langle \{s_{1}, s_{3}\}, \{h_{3}, h_{4}\}\rangle$, $\{h_{2}, h_{3}, h_{4}\}$ are the second-order neighbors of $h_{1}$ via the connections with $s_1$.
However, $h_{3}$ and $h_{4}$ do not appear with $h_{1}$ in the same prescription.
Thus there will be no edges between the $h_{1}$ and $h_{3}$ and between $h_{1}$ and $h_{4}$ in the synergy graphs.
On the other hand, it is obvious that the bipartite symptom-herb graph cannot be directly derived from the homogeneous synergy graphs.
In consequence, we conclude that the symptom-herb graph and synergy graphs contain their own characteristics and can complement each other to learn more accurate node representations.

\subsubsection{Synergy Graphs Construction}
Generally, the herb and symptom synergy patterns can be reflected by the high co-occurrence frequencies. 
Taking the construction of herb-herb graph as an example, we first compute the frequency of all herb-herb pairs in prescriptions: if herb $h_{m}$ and herb $h_{n}$ co-occur in the same $hc$, the frequency of pair $(h_{m},h_{n})$ is increased by 1. 
After obtaining the herb-herb frequency matrix, we manually set a threshold $x_{h}$ to filter the entries.  For pairs co-occurring more than $x_{h}$ times, the corresponding entries are set to 1, and 0 otherwise.   
It is formulated as follows,
$$ HH_{h_{m},h_{n}},HH_{h_{n},h_{m}}=\left\{
\begin{aligned}
1, &  \text{ if frequency (} h_{m}, h_{n} \text{)}  >  x_{h} ; \\
0, &  \text{ otherwise}, 
\end{aligned}
\right. 
$$
wherein $HH$ denotes the herb-herb graph. $x_{h}$ is the threshold for herb-herb pairs.
By referring to the above procedures, the symptom-symptom graph can be constructed as well.

\subsubsection{Information Propagation}
Given the constructed herb-herb graph $HH$ and symptom-symptom graph $SS$, we apply an one-layer graph convolution network to generate the symptom and herb embeddings:
\begin{equation} 
\begin{split}
&r_{s} = tanh (\sum_{k\in N^{SS}_{s}} e_{k} \cdot V_{s}),\\
&r_{h} = tanh (\sum_{q\in N^{HH}_{h}} e_{q} \cdot V_{h}),
\end{split}
\label{synergy_gcn}
\end{equation}
wherein $e_{k}$ and $e_{q}$ are initial embeddings for symptom $k$ and herb $q$ respectively.  $N^{SS}_{s}$ indicates the neighbor set of $s$ in $SS$.  $N^{HH}_{h}$ indicates the neighbor set of $h$ in $HH$.  $V_{s}$ and $V_{h}$ are weight parameters for $SS$ and  $HH$, respectively.
Through our local computation,
the averages of node degrees show that the symptom-herb graph is much denser than the synergy graphs, and the standard deviations verify that the degree distributions of synergy graphs are smoother than that of the symptom-herb graph. Considering that we need to fuse $b$ and $r$ lately, the sum aggregator is adopted for synergy graphs to make these two parts more balanced, which can benefit the training process to some extent.

From the view of herb recommendation task, $SS$ and $HH$ encode the synergy patterns in TCM, which further help improve the representation quality for symptoms and herbs.  Besides, introducing additional information helps relieve the data sparsity problem \cite{Ruan2019DiscoveringRF} of TCM prescriptions to some extent. 

\subsection{Information Fusion}
Up to now we have obtained two types of embeddings from Bipar-GCN and synergy graphs for each node.  We employ the simple addition operation to merge these embeddings,
\begin{equation} 
\begin{split}
&e_{s}^{*} = b_{s} + r_{s},\\
&e_{h}^{*} = b_{h} + r_{h},  
\end{split}
\end{equation}
wherein $e_{s}^{*}$ and $e_{h}^{*}$ are the merged embeddings for symptom $s$ and herb $h$, respectively.

To sum up, the above procedures clarify the proposed Multi-Graph Embedding Layer. 
It is a general architecture that can be used in other scenarios to model interactions between two types of objects.
For example, in the recommendation scenario, Bipar-GCN can be exploited to capture the intrinsic difference between users and items. The additional user-user graph can be the social relation graph among users. The item-item graph can be item relations linked by items' content attributes.

\subsection{Syndrome Induction}
As aforementioned,  syndrome induction plays an essential role in TCM clinical practice. 
Considering the ambiguity and complexity of syndrome induction,  in this work, we propose an MLP-based method to consider the implicit syndrome induction process, which can depict the nonlinear interaction among symptoms and generate an overall implicit syndrome representation. 

\begin{figure}[!tb]
    \centering
    \includegraphics[width=0.7\linewidth]{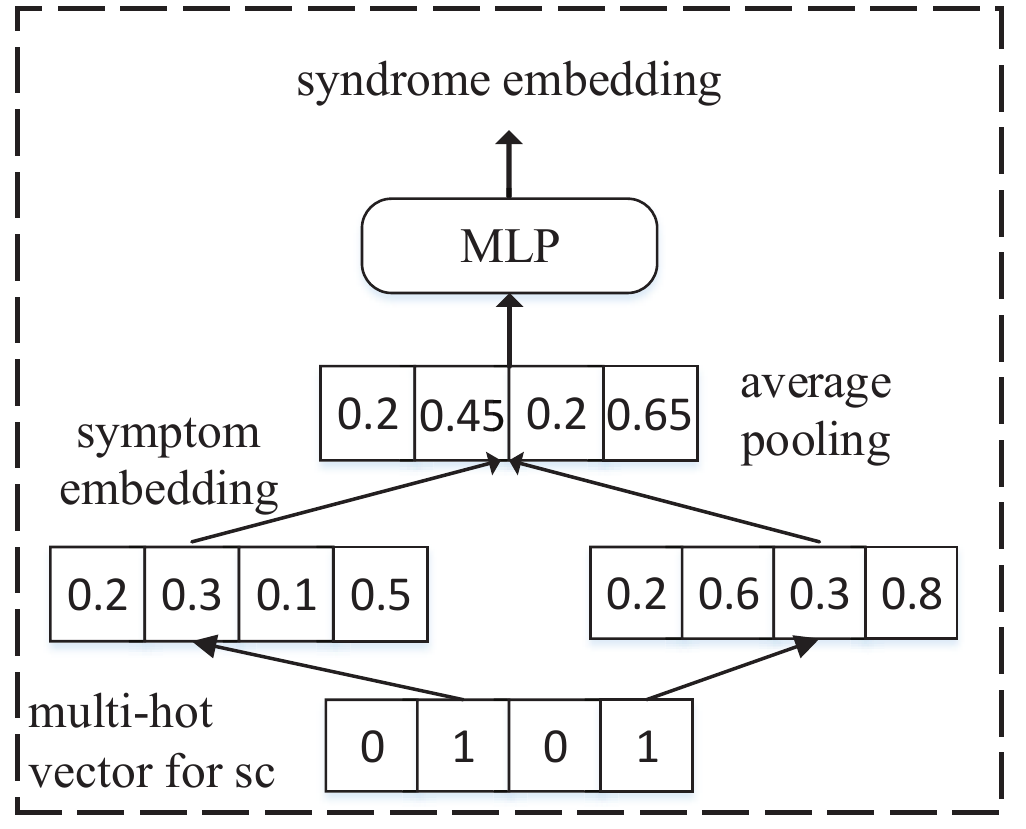}
    \caption{The MLP-based method for syndrome induction. }
    \label{fig:set_embedding}
\end{figure}

As Fig.~\ref{fig:set_embedding} shows, we feed all symptom embeddings in a symptom set into an MLP to induce the overall implicit syndrome representation.  Given a symptom set $sc$, first we represent it with a multi-hot vector.  In this vector, if $sc$ contains symptom $s$, the corresponding entry is set to 1, and 0 otherwise. 
Second we look up the embedding $e_{s}^{*}$ for each symptom $s$ in $sc$ and stack these vectors to build a matrix $e_{sc}\in R^{|sc|\times d}$, where $d$ is the dimension of the single symptom embedding. 
Third, to induce an overall representation from $e_{sc}$, average pooling ($Mean(\cdot)$) is utilized.  Further, considering the complexity of syndrome induction, we apply a single-layer MLP to transform the mean vector, which borrows the strength of nonlinearity in MLP to learn a more expressive syndrome representation. 
The above computation procedure is given as follows,
\begin{equation} 
e_{syndrome}(sc) = ReLU (\textbf{W}^{mlp}\cdot Mean (e_{sc})  + \textbf{b}^{mlp}) ,
\end{equation}
wherein $e_{syndrome}(sc)$ means the induced syndrome embedding for $sc$.

\subsection{ Training and Inference}

\begin{figure}[!tb]
    \centering
    \includegraphics[width=0.9\linewidth]{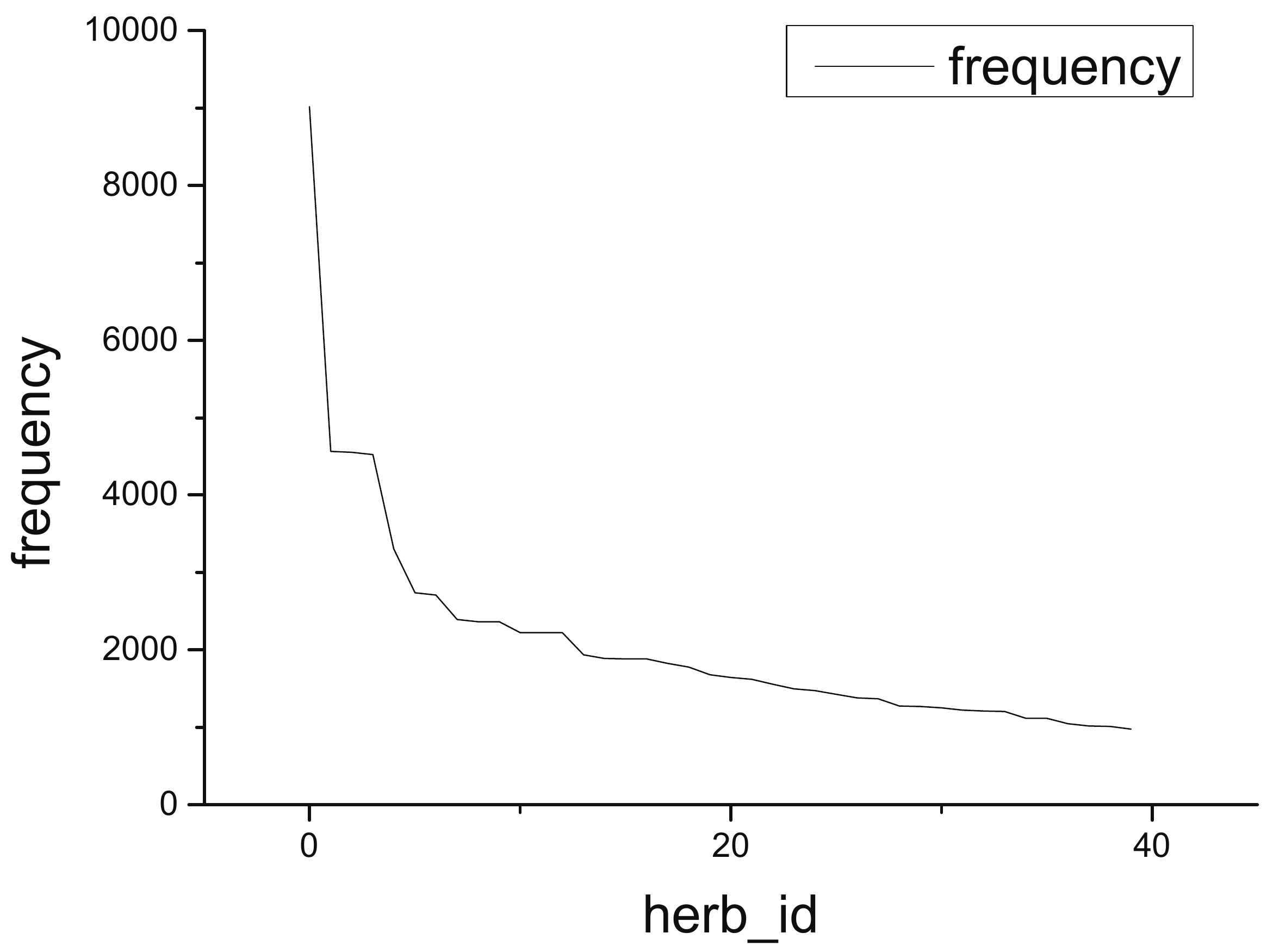}
    \caption{Frequency distribution of the top 40 most frequent herbs. }
    \label{fig:herb-frequency}
     
\end{figure}

In the herb recommendation scenario, given a symptom set, a herb set is generated to cure these symptoms.  For each prescription, we need to evaluate the distance between the recommended herb set and the ground truth herb set, which is similar to the multi-label classification task.  
As Fig. \ref{fig:herb-frequency} shows, the frequencies different herbs appear in prescriptions are imbalanced.  Therefore, we need to resolve the label imbalance problem.

Here, we use the following objective function (\ref{loss_function}) to characterize the above features in the herb recommendation scenario, where $e_{H}$ is the learned embedding matrix for the herb collection $H$.
 
\begin{small}
\begin{equation}
\begin{split}
Loss=\mathop{\arg\min}_{\theta}\sum_{ (sc, hc^{'}) \in P} &WMSE (hc^{'}, g(sc, H) )  + \lambda_{\Theta}||\Theta||_{2}^{2} \\
g(sc, H) &= e_{syndrome}(sc) \cdot e_{H}^{T}
\end{split}
\label{loss_function}
\end{equation}
\end{small}
Given the input  $sc$, the ground truth herb set $hc$ is represented as a  multi-hot vector $hc^{'}$ in dimension $|H|$.  $g (sc, H) $ is the output  probability vector for all herbs.   
$\lambda_{\Theta}$  controls the $L_{2}$ regularization strength to prevent overfitting. 
WMSE \cite{Hu2019Sets2SetsLF} is \textbf{weighted mean square loss} between   $hc^{'}$ and  $g (sc, H) $, which is defined as follows,
\begin{equation}
WMSE (hc^{'}, g(sc, H) ) =\sum_{i=1}^{|H|}w_{i} (hc^{'}_{i} - g (sc,H) _{i}) ^{2}
\label{eqa:wmse}
\end{equation}
The dimensions of $hc^{'}$ and $g (sc, H) $ are both  $|H|$.  $hc^{'}_{i}$ and $g (sc, H) _{i}$ indicate the i-th entries in vectors respectively. 
$w_{i}$ is the weight for herb $i$,
\begin{equation}
w_{i}=\frac{max_{k}freq (k) }{freq (i) }
\end{equation}
wherein $freq (i) $ is the frequency of herb $i$ appearing in prescriptions.   The adaptive weight setting is to balance the contribution of herbs with various frequencies.  As we can see, the more frequently herb $i$ appears, the lower its weight is.  We adopt Adam \cite{Kingma2014AdamAM} to optimize the prediction model and
update the model parameters in a mini-batch fashion.   
 
Some researches argue that there are some patterns among different labels that can be exploited to improve the performance in multi-label classification.   Zhang et al. \cite{Zhang2006MultilabelNN} introduce a regularization term to maximize the probability margin between the labels belonging to a set and the ones not belonging to the set.  However, the pair-wise margin is not reasonable in our scenario.  The detailed discussion is in the experiments part. 
 
\textbf{Inference}: Following the setting in  \cite{Hu2019Sets2SetsLF}, we also adopt the greedy strategy to generate the recommended herb set.  Specifically, we select the top $k$ herbs with the highest probabilities in $g (sc, H) $ as the recommended herb set for $sc$.

\section{Experiments}
\label{expe}

In this section, we evaluate our proposed \textbf{SMGCN} on the benchmark TCM dataset \cite{Yao2018ATM}.  There are several important questions to answer:

\textbf{RQ1}: Can our proposed model outperform the state-of-art herb recommendation approaches?

\textbf{RQ2}:  Can our proposed model outperform the state-of-the-art graph neural network-based recommendation approaches?

\textbf{RQ3}: How effective are our proposed components  (Bipar-GCN, Synergy Graph Encoding (SGE), and Syndrome Induction (SI))?

\textbf{RQ4}: How does our model performance react to different hyper-parameter settings (e.g.,  hidden layer dimension, depth of the GCN layers, and regularization strength)?

\textbf{RQ5}: Can our proposed SMGCN provide reasonable herb recommendation?

We first introduce the TCM data set, baselines, metrics, and experimental setup.  Then the experimental results are demonstrated in detail.  Last, we will discuss the influence of several critical hyperparameters. 

\setcounter{subsection}{0}
\subsection{Dataset}
\begin{figure}[!tb]
    \centering
    \includegraphics[width=1.0\linewidth]{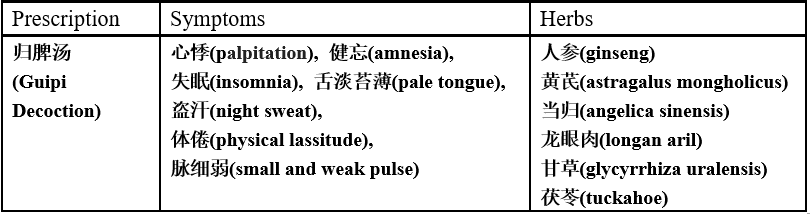}
    \caption{The prescription example. }
    \label{fig:data_example}
\end{figure}

To be consistent with work \cite{Wang2019AKG}, we conduct experiments on the benchmark  TCM data set \cite{Yao2018ATM}.  
The TCM data set contains 98,334 raw medical prescriptions and 33,765 processed medical prescriptions  (only consisting of symptoms and herbs).  
As Fig.~\ref{fig:data_example} shows, each prescription contains several symptoms and the corresponding herbs.  
Among 33,765 processed medical cases, Wang et al. \cite{Wang2019AKG} further select 26,360 prescriptions.  The 26,360 medical cases are divided into 22,917 for training and 3,443 for testing.  The statistics of the  experimental data
set is summarized in Tab.~\ref{tab:stat_data}. 
\begin{table}[!tb]
\caption{Statistics of the evaluation data sets}
\centering
\begin{tabular}{|c|c|c|c|}
\hline
Dataset & \#prescriptions & \#symptoms & \#herbs \\ \hline
All     & 26,360          & 360        & 753     \\ \hline
Train   & 22,917          & 360        & 753     \\ \hline
Test    & 3,443           & 254        & 558     \\ \hline
\end{tabular}
    \label{tab:stat_data}
\end{table}
\subsection{Evaluation}
Given a symptom set, our proposed model generates a herb set to relieve the symptoms.  To evaluate the performance of our approach, we adopt the following three measures commonly used in recommender systems.  For all prescriptions $ (sc,hc) $ in the test data set, they are defined by,

\begin{equation}
Precision@K = \frac{|Top (sc,K) \bigcap hc|}{K}
\end{equation}

\begin{equation}
Recall@K = \frac{|Top (sc,K) \bigcap hc|}{|hc|}
\end{equation}

\begin{equation}
NDCG@K = \frac{DCG@K}{IDCG@K}
\end{equation}
wherein $Top (sc,K) $ is the top K herbs with the highest prediction scores given $sc$. 
The $Precision@K$ score indicates the hit ratio of top-K herbs as true herbs. In the experiments, we use $Precision@5$ to decide the optimal parameters. 
The $Recall@K$ describes the coverage of true herbs as a result in top-K recommendation. 
The $NDCG@K$  (Normalized Discounted Cumulative Gain)  accounts for the position of the hit herbs in the recommended list.  If the hit herb ranks higher in the list, it gains a larger score. 
We truncate the ranked list at 20 for all three measures and report the average metrics for all prescriptions in the test set. 

\subsection{Baselines}
We adopt the following approaches for comparison.

\textbf{Topic model}
\begin{itemize}
    \item HC-KGETM \cite{Wang2019AKG}: It integrates the TransE \cite{Bordes2013TranslatingEF} embeddings obtained from a TCM knowledge graph into the topic model, to consider not only the co-occurrence information in TCM prescriptions but also comprehensive semantic relatedness of symptoms and herbs from the knowledge graph. 
\end{itemize}

\textbf{Graph neural network-based Models}
\begin{itemize}
    \item GC-MC \cite{berg2017graph}: This model leverages GCN \cite{Kipf2016SemiSupervisedCW} to obtain the representations of users and items.  To be consistent with the original work,  we set one graph convolution layer in the experiment, and the hidden dimension equals the embedding size.  
    \item PinSage \cite{Ying2018GraphCN}:  PinSage is an industrial application of GraphSAGE \cite{Hamilton2017InductiveRL} on item-item graph.  In our setting, we apply it on the symptom-herb
interaction graph.  Specifically, we adopt two graph convolution
layers following \cite{Ying2018GraphCN}, and the hidden dimension is the same as the embedding size. 
    \item NGCF \cite{wang2019neural}:  NGCF  is the state-of-the-art graph-based collaborative filtering method.  It explicitly constructs a bipartite user-item graph to model the high-order connectivity and obtain more expressive representations for users and items. 

\end{itemize}

\textbf{Our proposed models}    
\begin{itemize}
 \item HeteGCN: It is our proposed baseline which is built based on the heterogeneous graph-based GCN \cite{zhang2019heterogeneous}.
    We integrate the symptom-herb graph, herb-herb graph, and symptom-symptom graph into one heterogeneous graph. For each node, there are two types of neighbors, symptom neighbors and herb neighbors. We apply the type-based attention mechanism to perform message construction. For symptom $s$, the one-hop neighbor message is in (\ref{attention_1}) and (\ref{attention_2}), where $tp$ = $\{symptom, herb\}$ denotes the neighbor type set, $m$ is defined in (\ref{info}), and $||$ indicates the concatenation operation. $W^{att}$ and $z$ are the attention network parameters. The information propagation is the same as (\ref{b_s_1}). To notice that, symptom and herb nodes share the same network parameters. The formulas for herb nodes are similar. HeteGCN adopts the average pooling to do syndrome induction, and multi-label loss is defined similar to (\ref{loss_function}). 
    The depth of GCN is set to 1 with hidden dimension of 128 for better performance.
    
 \begin{small}
\begin{equation} 
 b_{N_{s}}^{0} = tanh (\sum_{t \in tp} \alpha^{t}\frac{1}{|N_{s}^{t}|} \sum_{n\in N_{s}^{t}}m_{n}^{0})
\label{attention_1}
\end{equation}
\end{small}
\begin{small}
\begin{equation} 
\alpha^{t} = \frac{exp(z^{T}  ReLU(W^{att}\cdot(e_{s}  || \frac{1}{|N_{s}^{t}|} \sum\limits_{n\in N_{s}^{t}}m_{n}^{0})))}{\sum_{t' \in tp}exp(z^{T}ReLU(W^{att}\cdot(e_{s} || \frac{1}{|N_{s}^{t'}|} \sum\limits_{n\in N_{s}^{t'}}m_{n}^{0})))}
\label{attention_2}
\end{equation}
\end{small}

    \item SMGCN: The proposed approach learns multiple graphs (i.e., the symptom-herb bipartite graph, symptom-symptom graph, and herb-herb graph), and performs graph convolution on them to describe the complex relations between symptoms and herbs from TCM. In the prediction layer, we design an MLP-based method to induce the overall implicit syndrome representation for each symptom set. As a result, it is significantly different compared with existing herb recommendation methods.
\end{itemize}

\subsection{Parameter Settings}

We implement our approach and the comparative methods using Tensorflow. 
For the topic model HC-KGETM, we follow the parameter settings in  \cite{Wang2019AKG}. 
Grid search is conducted to search the optimal learning rate $lr$, the regularization coefficient $\lambda_\Theta$ and the dropout ratio.
Specifically, $lr$ is varied in $\{10^{-5}, 10^{-4}, 10^{-3}\}$, $\lambda_\Theta$ is tuned in $\{0, 10^{-6}, 10^{-5}, 10^{-4}, 10^{-3}\}$, and the dropout rate is searched in $\{0, 0.1, ..., 0.8\}$.
We use Xavier initializer \cite{Glorot2010UnderstandingTD} and Adam optimizer \cite{Kingma2014AdamAM}  to train models with the batch size of 1024.   

For graph neural network baselines,  the embedding size and the latent dimension are both set to 64.   For our proposed SMGCN and HeteGCN, the embedding size is fixed to 64, and the dimension of the first output layer is 128.  The last layer dimension is searched in $\{64, 128, 256, 512\}$.  The GCN layer depth is tuned in $\{1, 2, 3\}$.  The optimal parameter settings are summarized in Tab.~\ref{tab:paras}. 
Without specification, the following performances of our SMGCN model are with 2 GCN layers and the last layer dimension of 256. 

\begin{table}[!tb]
\caption{Optimal parameters of comparative models}
\centering
\begin{tabular}{|l|l|}
\hline
Approaches  & Best parameter settings                                                                                                      \\ \hline
HC-KGETM    & \begin{tabular}[c]{@{}l@{}}$\alpha$ = 0.05  $\beta_{s}$ = $\beta_{h}$ = 0.01  $\gamma$ = 1\end{tabular}            \\ \hline
GC-MC       & \begin{tabular}[c]{@{}l@{}}lr = 9e-4 dropout = 0.0 $\lambda$ = 1e-6  
\end{tabular}                                              \\ \hline
PinSage     & \begin{tabular}[c]{@{}l@{}}lr = 9e-4 dropout = 0.0 $\lambda$ = 1e-3 
\end{tabular}                                              \\ \hline
NGCF        & \begin{tabular}[c]{@{}l@{}}lr = 3e-3 dropout = 0.0 $\lambda$ = 1e-5 
\end{tabular}                                              \\ \hline
HeteGCN        & \begin{tabular}[c]{@{}l@{}}lr = 3e-3 dropout = 0.0 $\lambda$ = 1e-3 \\  $x_{s}$ =5 $x_{h}$=40 \end{tabular} \\ \hline
SMGCN & \begin{tabular}[c]{@{}l@{}}lr = 2e-4 dropout = 0.0 $\lambda$ = 7e-3 \\  $x_{s}$ =5 $x_{h}$=40
\end{tabular} \\ \hline
\end{tabular}

    \label{tab:paras}
\end{table}

\subsection{Performance Comparison}
In this part, we firstly demonstrate the overall results among different
methods, with their optimal parameter settings.  Next, we conduct some ablation analysis to verify the effectiveness of different model components.  Then we discuss the influence of hyperparameters in detail. 

\subsubsection{Overall Result} (RQ1\&RQ2)

Tab.~\ref{tab:performance} demonstrates the overall performances.  To notice that,  the original graph neural network-based baselines do not apply Syndrome Induction (SI) and multi-label loss functions.  For a fair comparison,   we modify GC-MC, PinSage and NGCF by adding the SI part and employing multi-label loss function defined in (\ref{loss_function}). We can observe that:

\begin{table*}[!ht]
\caption{The overall performance comparison. HC-KGETM utilizes  log-loss but without SI. HeteGCN utilizes multi-label loss but without SI. The other models are with SI and adopt multi-label loss.  The second best results are underlined.  p@k and r@k are short for precision@k and recall@k}
\centering
\resizebox{0.97\textwidth}{!}{
\begin{tabular}{|c|c|c|c|c|c|c|c|c|c|}
\hline
Approaches                                 & p@5                                      & p@10                                      & p@20                                      & r@5                                       & r@10                                      & r@20                                      & ndcg@5                                    & ndcg@10                                   & ndcg@20                                   \\ \hline
HC-KGETM                                   & 0.2783                                  & 0.2197                                   & 0.1626                                   & 0.1959                                   & 0.3072                                   & 0.4523                                   & 0.3717                                   & 0.4491                                   & 0.5501                                   \\ \hline
GC-MC                                      & 0.2788                                  & 0.2223                                   & 0.1647                                   & 0.1933                                   & 0.3100                                   & 0.4553                                   & 0.3765                                   & 0.4568                                   & 0.5610                                   \\ \hline
PinSage              & {0.2841}    & {0.2236}    & {0.1650}    & {0.1995}    & {0.3135}    & {0.4567}    & {\underline{0.3841}}    & {0.4613}    & {0.5647}    \\ \hline
NGCF                                       & 0.2787                                  & 0.2219                                   & 0.1634                                   & 0.1933                                   & 0.3085                                   & 0.4505                                   & 0.3790                                   & 0.4571                                   & 0.5599                                   \\ \hline
HeteGCN                                   & \underline{0.2864}                                   & \underline{0.2268}                                    & \underline{0.1676}                                  & \underline{0.2018}                                   & \underline{0.3192}                                    & \underline{0.4667}                                    & 0.3837                                   & \underline{0.4620}                                  & \underline{0.5665}                              \\ \hline
\textbf{SMGCN} & \textbf{0.2928} & \textbf{0.2295} & \textbf{0.1683} & \textbf{0.2076} & \textbf{0.3245} & \textbf{0.4689} & \textbf{0.3923} & \textbf{0.4687} & \textbf{0.5716} \\ \hline
 \%Improv.  by HC-KGETM  & 5.22\%          & 4.44\%          & 3.52\%          & 5.95\%          & 5.63\%          & 3.67\%          & 5.55\%          & 4.36\%          & 3.90\%          \\ \hline
  \%Improv.  by PinSage   & 3.09\%          & 2.61\%          & 2.02\%          & 4.02\%          & 3.49\%          & 2.68\%          & 2.13\%          & 1.60\%          & 1.23\%          \\ \hline
 \%Improv.  by HeteGCN  & 2.24\%                                 & 1.17\%                                    & 0.44\%                                    & 2.87\%                                   & 1.66\%                                  & 0.46\%                                   & 2.24\%                                  & 1.45\%                                    & 0.90\%                                    \\ \hline
\end{tabular}
}
    \label{tab:performance}
\end{table*}

\begin{itemize}
\item  
Our proposed SMGCN  performs the best among the comparative approaches.  Specifically, SMGCN outperforms the topic-model HC-KGETM in terms of p@5 by $5. 22\%$, r@5 by $5. 95\%$ and ndcg@5 by $5. 55\%$.  Besides, as for the strongest baseline HeteGCN, SMGCN outperforms it in terms of p@5 by $2.24\%$, r@5 by $2.87\%$, and ndcg@5 by $2.24\%$. For the second best baseline PinSage, SMGCN surpasses it in terms of p@5 by $3.09\%$, r@5 by $4.02\%$, and ndcg@5 by $2.13\%$.

\item
HC-KGETM almost performs the worst for all metrics.  The reasons may contain two aspects: 1)  at the interaction-modeling stage, it only ranks the candidate herbs based on each single symptom and ignores the symptom set information;
 2)  at the embedding learning step,  it adopts TransE \cite{Bordes2013TranslatingEF} to capture the information in a TCM knowledge graph.  Compared to the translation-based graph embedding method, the graph neural networks are superior in explicitly exploiting the high-order connectivity.  

\item 
 Among GC-MC, PinSage, and NGCF, NGCF performs the worst, and PinSage performs the best.  Comparing GC-MC with NGCF, GC-MC performs slightly better than NGCF.   Considering that GC-MC only utilizes the first-order neighbors, the multiple graph convolution layers of NGCF may cause overfitting and hurt the performance. 
Further,  PinSage, GC-MC, and NGCF have various propagation functions: PinSage concatenates representations of the target node and the neighbor nodes, GC-MC sums these two representations, and NGCF additionally integrates the element-wise product part of the target node and the neighbor node when constructing messages.  It seems that the concatenation operation is more effective in capturing the rich relations in prescriptions, compared with the element-wise product or sum-up operations. 

\item
HeteGCN outperforms PinSage, which shows that additionally integrating the herb-herb and symptom-symptom concurrent relations can introduce more information. However, SMGCN is still superior to HeteGCN, which verifies that the Multi-Graph GCN framework can learn a more flexible and expressive model to some extent compared with the unified heterogeneous-graph based GCN, and the choice of MLP is appropriate to depict the complex syndrome induction process.

\end{itemize}
 
\subsubsection{Ablation Analysis} (RQ3)

To better understand our proposed SMGCN model, we split the whole model into three components: Bipar-GCN,  SGE, and Syndrome Induction (SI)  to evaluate their contribution to the unified herb recommender system, respectively. 
To notice that, in Bipar-GCN, we only use average pooling to do syndrome induction for each symptom set.  In the SI part, we adopt the average pooling followed by an MLP transformation. 
Tab.~\ref{tab:part_experiment} shows the performance of the ablation analysis.  Here the output embedding size is set to 256, and the graph convolution layer is set to 2.  Among the submodels, Bipar-GCN and Bipar-GCN w/ SI do not contain the synergy graphs. Thus instead of the heterogeneous graph-based HeteGCN, we adopt the simpler baseline PinSage to be compared with all the submodels. We have the following observations:

\begin{table}[!tb]
\caption{Performance of different submodels }
\centering
\resizebox{0.45\textwidth}{!}{
\begin{tabular}{|c|c|c|c|}
\hline
Submodels                                & p@5    & r@5    & ndcg@5 \\ \hline
PinSage                    & 0.2841 & 0.1995 & 0.3841 \\ \hline
Bipar-GCN                                & 0.2859 & 0.2003 & 0.3820 \\ \hline
Bipar-GCN w/ SGE            & 0.2916 & 0.2064 & 0.3900 \\ \hline
Bipar-GCN w/ SI  & 0.2914 & 0.2060 & 0.3885 \\ \hline

SMGCN & \textbf{0.2928} & \textbf{0.2076} & \textbf{0.3923} \\ \hline
\end{tabular}}
    \label{tab:part_experiment}
\end{table}

\begin{itemize}
\item  From a whole perspective, all the three components of our proposed model, i.e., Bipar-GCN, SGE, and SI, are verified to be effective for their better performance in comparison. 

\item Comparing Bipar-GCN with Bipar-GCN w/ SI, it is observed that the choice of  MLP is superior to only employing average pooling, which verifies that the nonlinear transformation in MLP helps model the complex relations among symptoms and  further generate a high-quality implicit syndrome representation.

\item For both Bipar-GCN and Bipar-GCN w/ SI, integrating Synergy Graph Encoding (SGE) leads the further improvement, which shows that the architecture of multiple graphs in the embedding learning layer not only is beneficial for learning more expressive representations but also assist in predicting herbs. 

\item SMGCN, the combination of Bipar-GCN, SGE, and SI, achieves the best performance, indicating that modeling the nonlinearity in the syndrome inducing process and unifying complex relations through multiple graphs is effective in the herb recommendation scenario. 
\end{itemize}

\subsubsection{Influence of Hyperparameters} (RQ4) 

In this part, we will discuss the key factors in detail.

\begin{itemize}
\item Effect of Layer Numbers
\end{itemize}

To explore whether our proposed model can benefit from a larger number of embedding propagation layers, we tune
the number of GCN layers on the submodel  Bipar-GCN w/ SI, which is varied in $\{1, 2, 3\}$. The dimension of the last layer is set to 256.   We have the following observations from Tab.~\ref{tab:depth}:

\begin{itemize}
\item[-]  Our proposed Bipar-GCN w/ SI is not very sensitive to the depth of propagation layers. The two-layer model performs marginally better compared to one-layer's performance.

\item[-] When further increasing the layer number to three,  it seems that the performance drops a little compared to one layer.  The reason may be overfitting caused by large propagation depth. 

\item[-] When varying the depth of propagation layers, our Bipar-GCN w/ SI consistently outperforms the  strongest baseline HeteGCN.  
It again verifies the effectiveness of the SI part, empirically showing that the nonlinearity of MLP  can help depict the complex syndrome induction process.
 
\end{itemize}

\begin{table}[!tb]
\caption{Effect of layer numbers on  Bipar-GCN w/ SI}
\centering
\setlength{\tabcolsep}{1mm}{
\begin{tabular}{|c|c|c|c|c|c|c|}
\hline
depth & p@5    & p@20   & r@5    & r@20   & ndcg@5 & ndcg@20 \\ \hline
1     & 0.2898 & 0.1688 & 0.2044 & \textbf{0.4702} & 0.3864 & 0.5684  \\ \hline
2     & \textbf{0.2914} & \textbf{0.1690} & \textbf{0.2060} & 0.4695 & \textbf{0.3885} & \textbf{0.5699}  \\ \hline
3     & 0.2882 & 0.1684 & 0.2030 & 0.4684 & 0.3869 & 0.5693  \\ \hline
\end{tabular}}
    \label{tab:depth}
\end{table}
 
\begin{itemize}
\item Effect of Final Embedding Dimension
\end{itemize}

The dimension of the embedding
layer can influence the performance
a lot.  We conduct the experiments on our proposed SMGCN approach, and the depth of embedding propagation is set to 2. 
Tab.~\ref{tab:dimension}  shows the experimental
results according to various dimensions of the last output layer. 
With the output dimension increasing, there is a consistent improvement with a larger embedding dimension until dimension to be 256.  When the dimension rises at 512, the performance drops slightly but is still superior to the second strongest baseline PinSage.  However, when the dimension drops to 64, our model underperforms PinSage in terms of r@20 and ndcg@20 slightly. 
This observation denotes that our proposed model depends on a reasonably large embedding dimension to have sufficient flexibility for constructing
useful embeddings.  

\begin{table}[!tb]
\caption{Effect of last layer dimensions on SMGCN }
\centering
\setlength{\tabcolsep}{1mm}{
\begin{tabular}{|c|c|c|c|c|c|c|}
\hline
dimension    & p@5             & p@20            & r@5             & r@20            & ndcg@5          & ndcg@20         \\ \hline
64           & 0.2857          & 0.1651          & 0.1999          & 0.4554          & 0.3847          & 0.5627          \\ \hline
128          & 0.2882          & 0.1670          & 0.2018          & 0.4631          & 0.3853          & 0.5660          \\ \hline
256 & \textbf{0.2928} & \textbf{0.1683} & \textbf{0.2076} & \textbf{0.4689} & 0.3923          & \textbf{0.5716} \\ \hline
512 & 0.2922          & 0.1673          & 0.2068          & 0.4632          & \textbf{0.3930} & 0.5700          \\ \hline
\end{tabular}}

    \label{tab:dimension}
\end{table}

\begin{itemize}
   \item Effect of Frequency Thresholds in Synergy Graphs
\end{itemize}

\begin{figure}[!tb]
	  \subfloat[precision@5]{
       \includegraphics[width=1in]{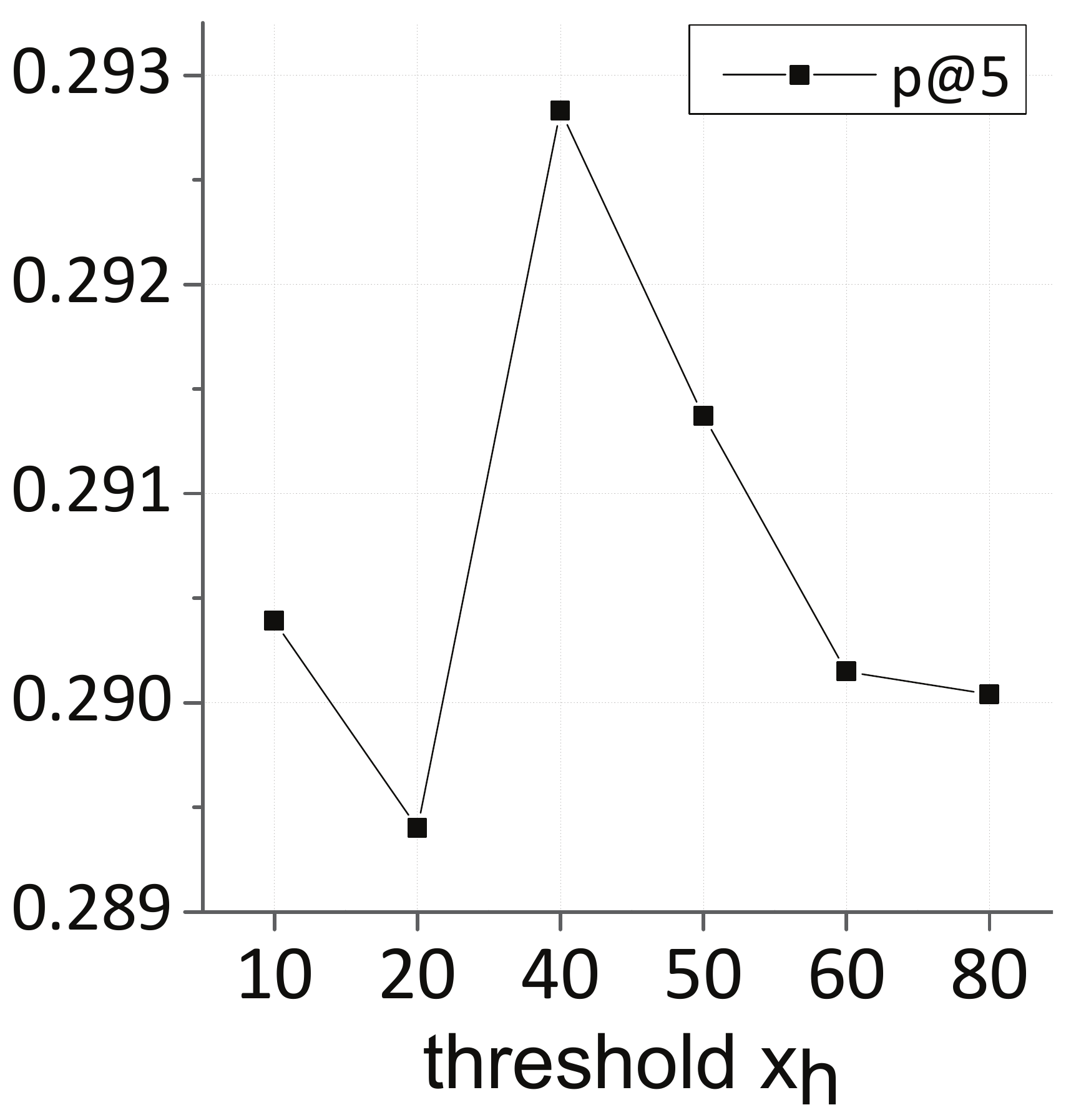}}
     \quad
	  \subfloat[recall@5]{
        \includegraphics[width=1in]{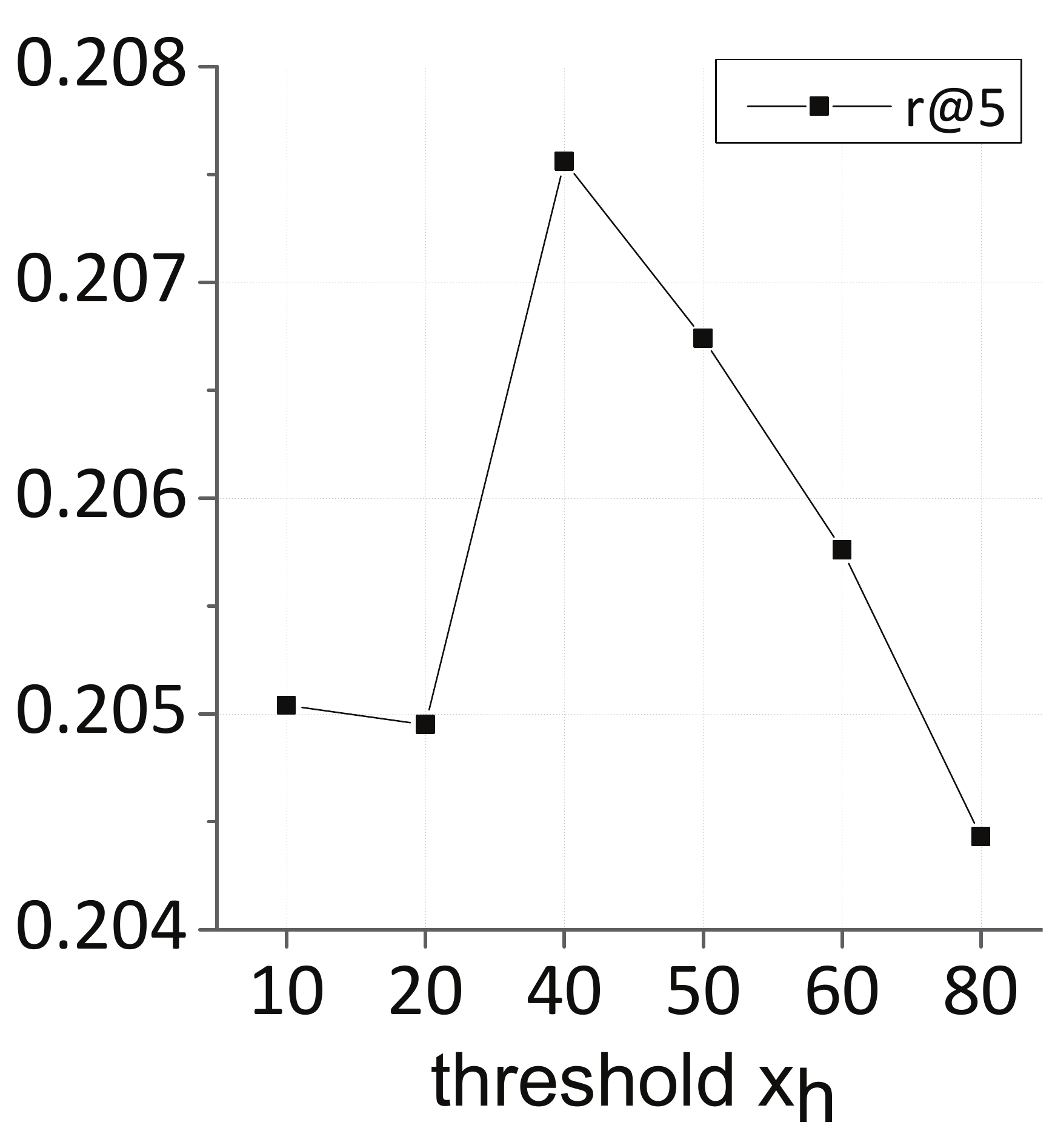}}
      \quad
	  \subfloat[ndcg@5]{
        \includegraphics[width=1in]{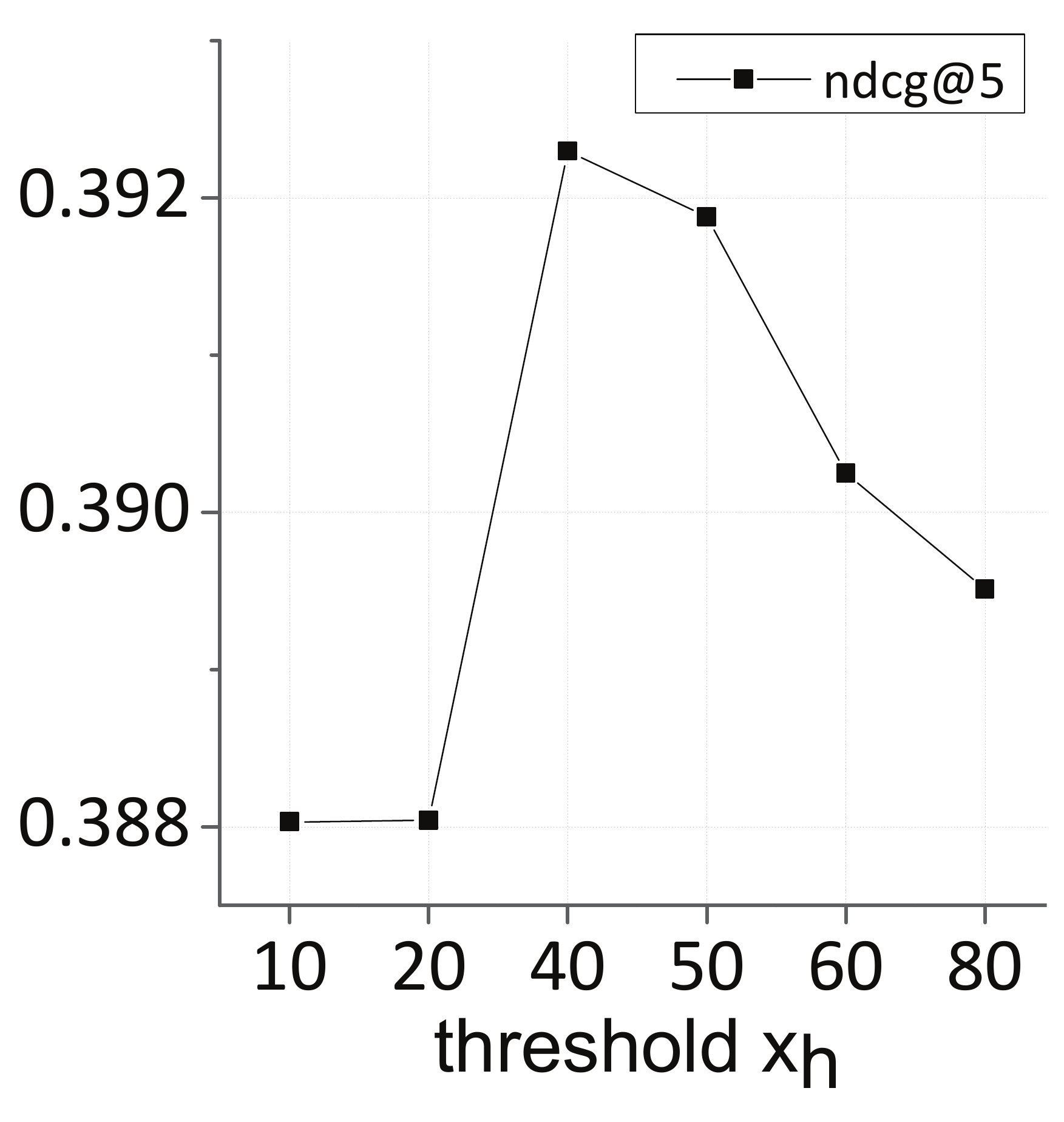}}
	  \caption{Performance for different thresholds on SMGCN.}
	  \label{figure:threshold} 
\end{figure}

The Synergy  Graph Encoding (SGE) component containing symptom-symptom graph and herb-herb graph contributes a lot to our proposed SMGCN.  These two graphs are used to reflect the concurrency patterns between herb-herb pairs and symptom-symptom pairs, which play an important role in TCM theory.   There are two hyperparameters controlling the construction of synergy graphs, threshold $x_{h}$ for herb-herb co-occurrence and $x_{s}$ for symptom-symptom co-occurrence.  For instance, if the symptom-symptom pair $ (s_{m},s_{n}) $ occurs in prescriptions more than $x_{s}$ times, then edge $ (s_{m},s_{n}) $ is added into the symptom-symptom graph.  We fix $x_{s}$ to 5 and tune $x_{h}$ varied in $\{10, 20, 40, 50, 60, 80\}$.  Fig.~\ref{figure:threshold} shows the experimental results for different thresholds.
We show the metrics in terms of topk=5, and the slightly better performance is achieved at $x_{h}$=$40$.  When the threshold is low, the herb-herb graph is relatively dense, but it may contain some noise.  As the threshold increases, the graph becomes sparse, and some useful information may be filtered.  Therefore, finding an appropriate threshold seems to affect the construction of synergy graphs.

\begin{itemize}
\item Effect of Regularization
\end{itemize}

\begin{figure}[!tb]
	  \subfloat[precision@5]{
       \includegraphics[width=1in]{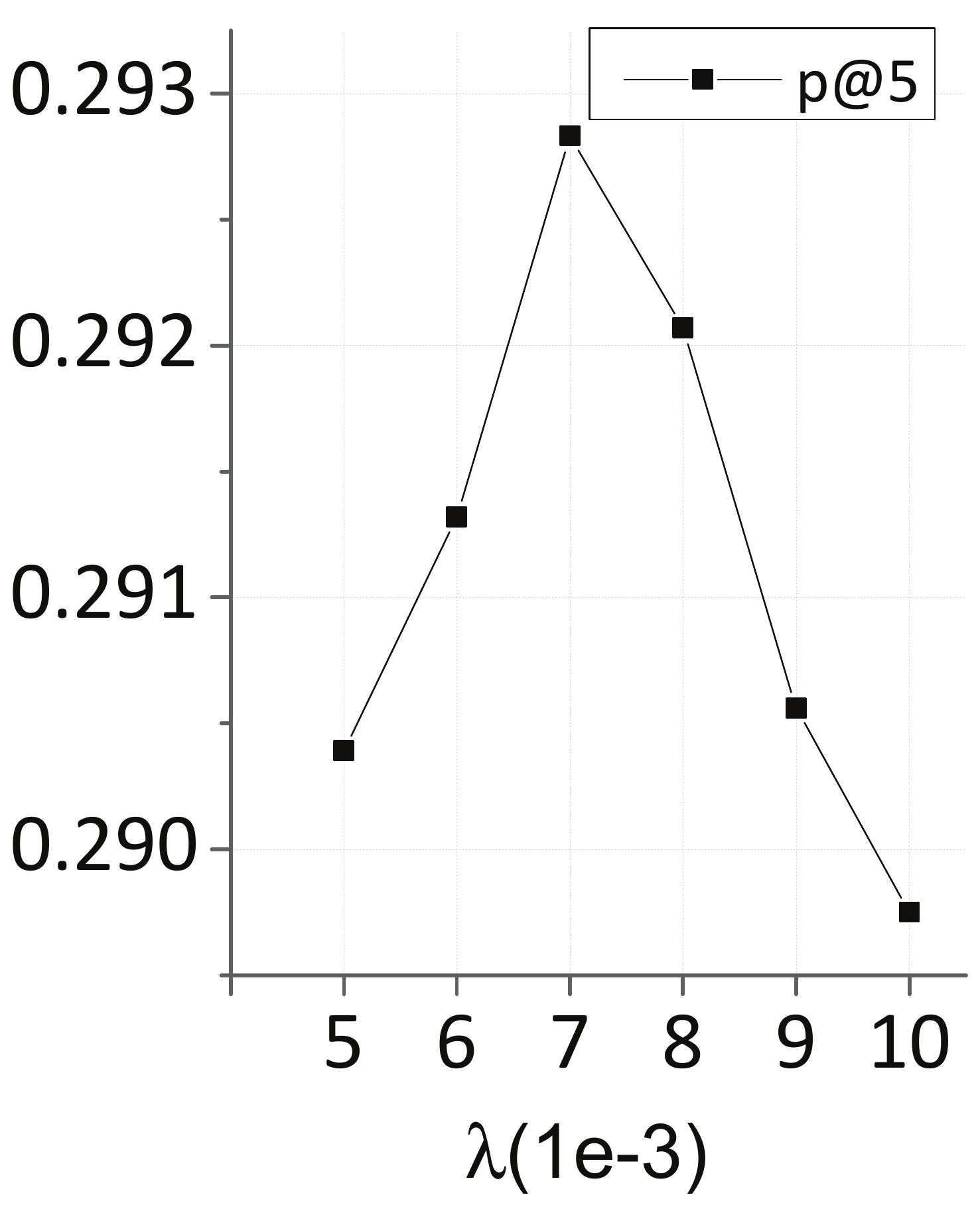}}
     \quad
	  \subfloat[recall@5]{
        \includegraphics[width=1in]{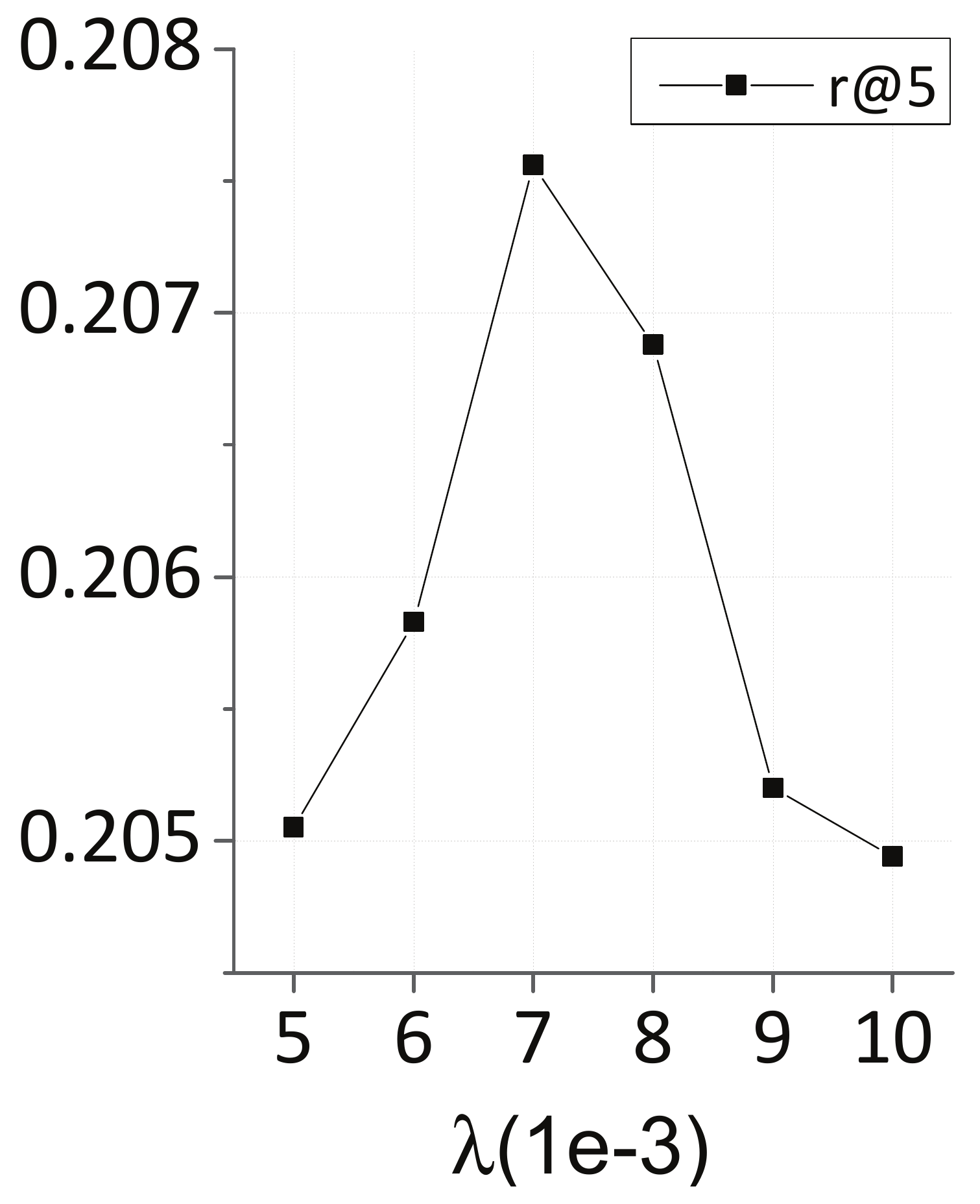}}
      \quad
	  \subfloat[ndcg@5]{
        \includegraphics[width=1in]{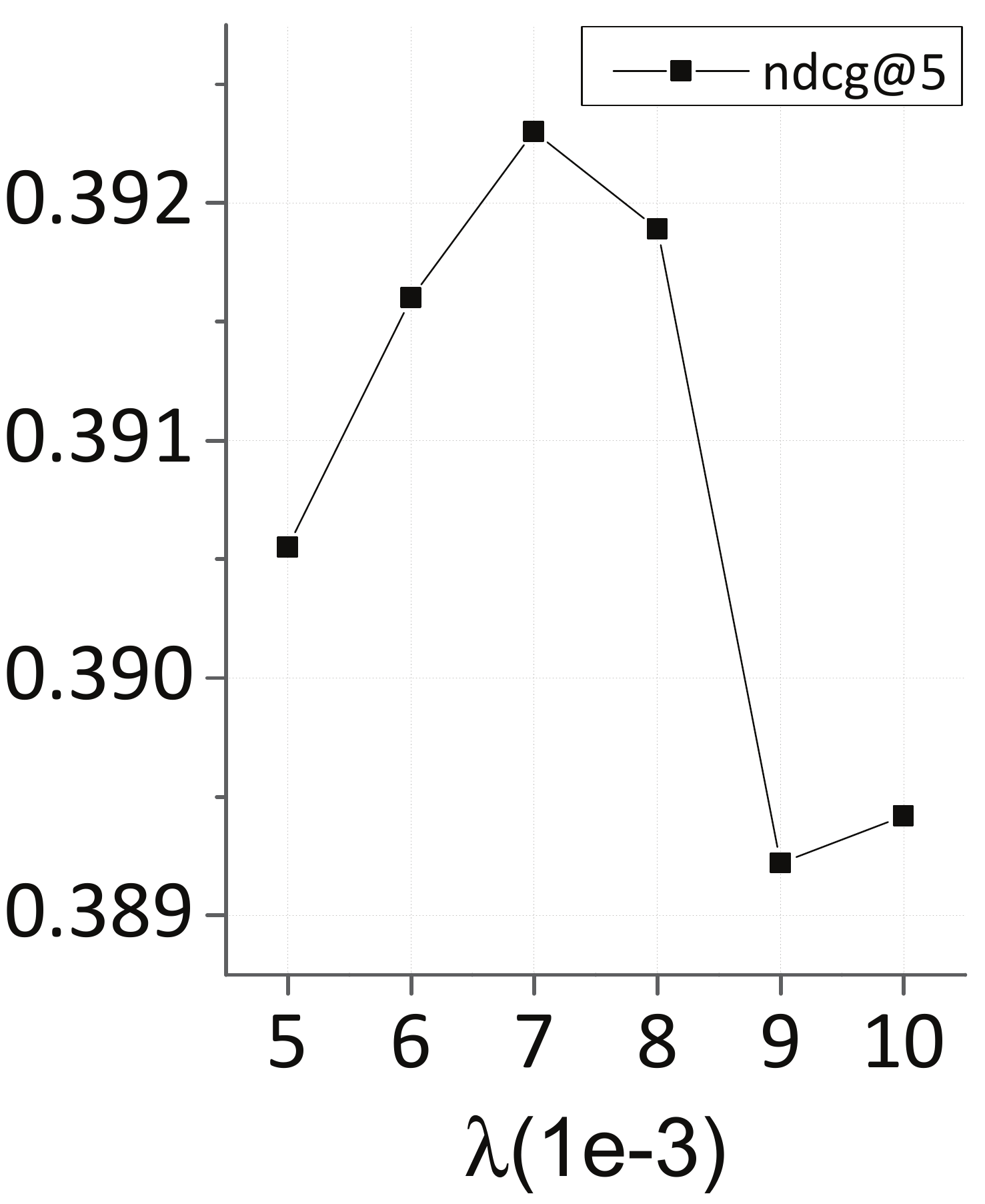}}
	  \caption{Performance for different $\lambda$ on SMGCN.}
	  \label{figure:reg} 
\end{figure}

\begin{figure}[!tb]
	  \subfloat[precision@5]{
       \includegraphics[width=1in]{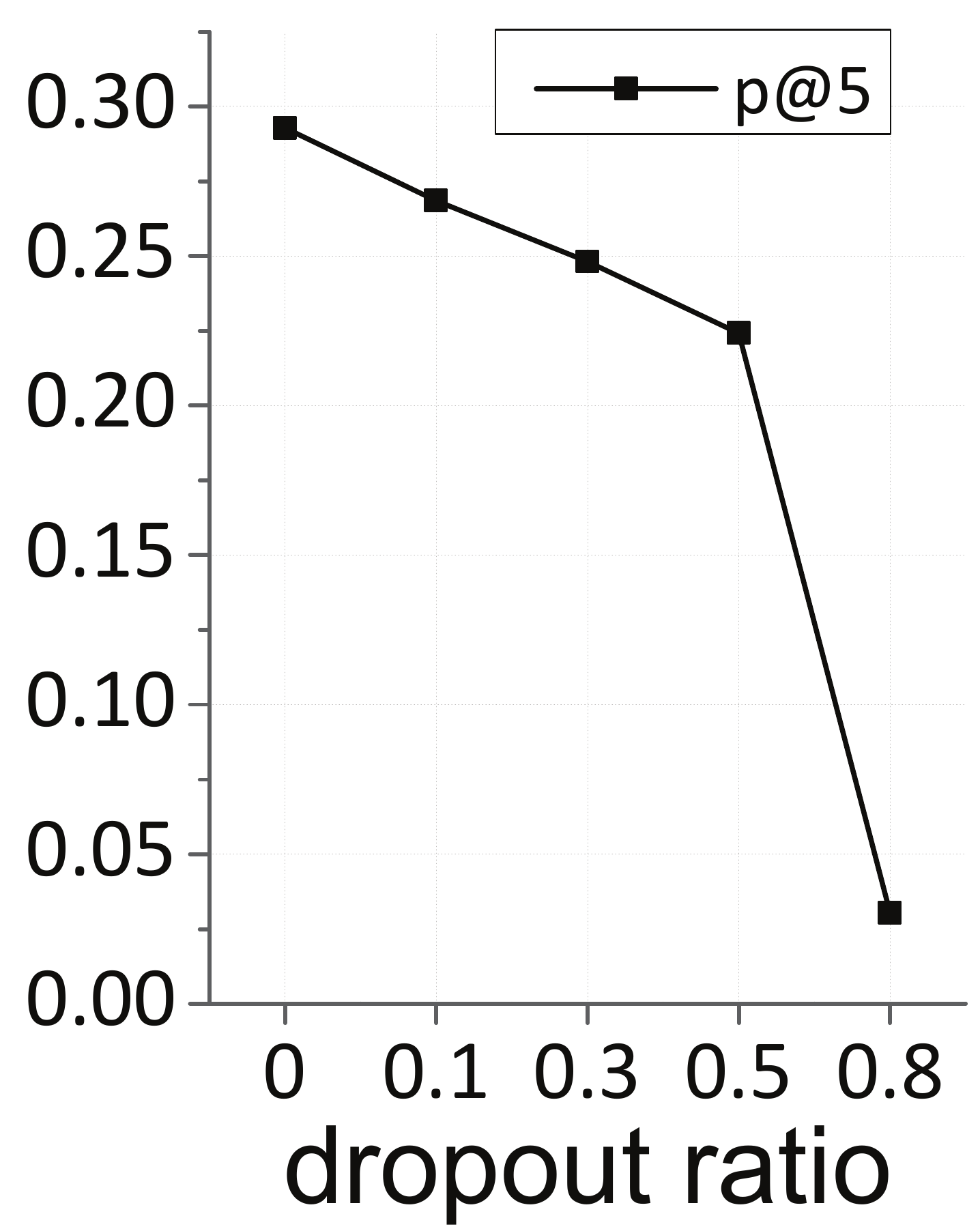}}
     \quad
	  \subfloat[recall@5]{
        \includegraphics[width=1in]{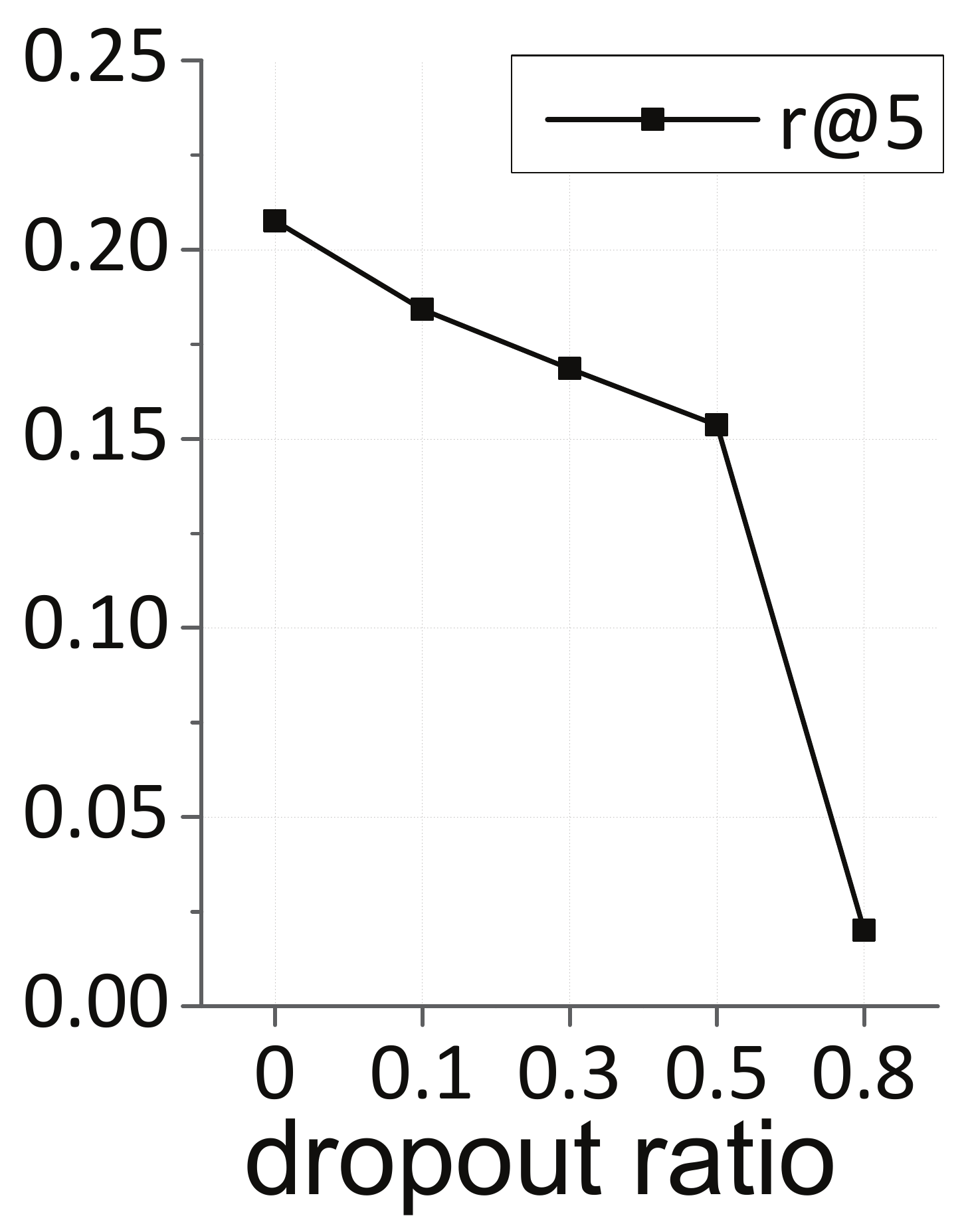}}
      \quad
	  \subfloat[ndcg@5]{
        \includegraphics[width=1in]{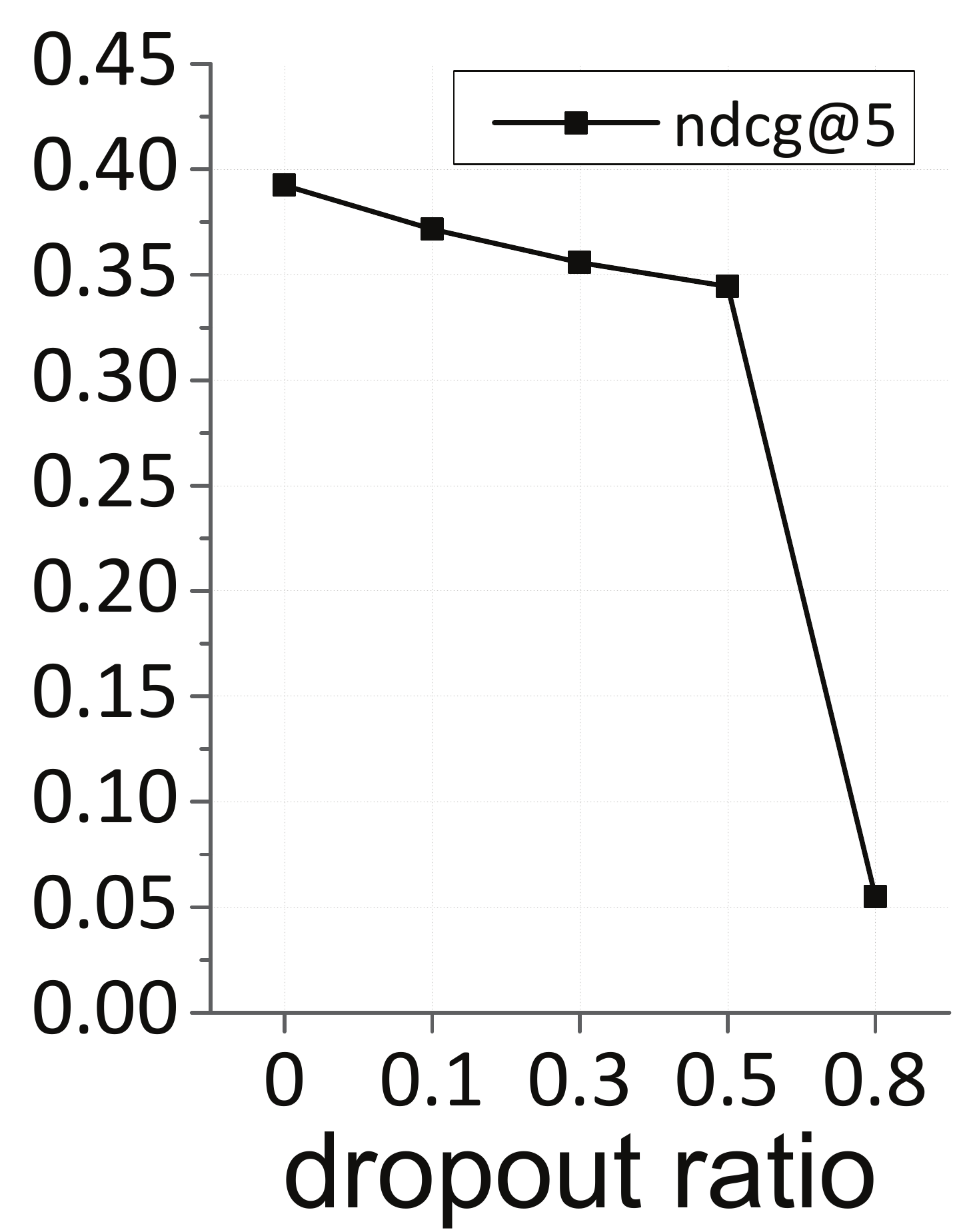}}
	  \caption{Performance for different dropout ratios on SMGCN.}
	  \label{figure:dropout} 
\end{figure}

Due to the strong expressiveness
of neural networks, it is easy to overfit the training data.  The typical approaches to prevent overfitting contain regularization term
and the dropout of neurons.  In our setting, $\lambda$ controls the regularization
strength on parameters, and the dropout ratio controls the ratio of removed neurons in the training process.   Fig.~\ref{figure:reg} demonstrates the influence of $\lambda$ and Fig.~\ref{figure:dropout} depicts the influence of the dropout ratio, where the dimension is set to 256, and the depth is set to 2.  
From Fig.~\ref{figure:reg}, we observe that our model achieves slightly better performance when $\lambda$ equals 7e-3.  Larger $\lambda$ might result in under-fitting and hurt the performance.  Smaller $\lambda$ might be weak to prevent the overfitting trend in the training process. 
 As for the dropout technique, instead of dropping out some
nodes completely with a certain probability, we only employ message dropout on the aggregated neighborhood embeddings, making our model more robust against the presence or absence of single edges. 
It can be observed that the performance drops with the increasing dropout ratio, which indicates that the above regularization term is sufficient enough to prevent the overfitting trend. 

\begin{itemize}
\item Effect of Loss Function
\end{itemize}

In Tab.~\ref{tab:performance},  we align GNN based baselines (i.e., GC-MC, PinSage, and NGCF) with our proposed SMGCN by adding the SI component and employing multi-label loss on them.  Therefore, the performance comparison only verifies the effectiveness of the embedding learning layer in our model.  We are also curious about the effectiveness of the different embedding layer and prediction layer combinations.  
We select NGCF from the baselines as a representative method, and the comparative loss function is the common-used pair-wise BPR \cite{Rendle2009BPRBP}.  The experimental results are summarized in Tab.~\ref{tab:loss_function}.   
As for BPR loss, Bipar-GCN w/ SI performs better.  For multi-label loss, Bipar-GCN w/ SI is superior in all the metrics.  It verifies that separately learning symptom and herb representations can help obtain more expressive embeddings.  
Besides, multi-label loss also outperforms BPR loss, which tells that multi-label loss is more appropriate for herb-recommendation task than BPR loss. We will give the reasons in detail. In a TCM prescription, the herb set is generated according to herb compatibility rules, which heavily depend on the TCM doctors' individuals experiences.  For the same symptom set, there may be multiple herb sets as the remedy.  Therefore, when herb A occurs in a prescription, it does not mean that A is more appropriate than every missing herb B.  It just indicates that herb B is not reasonable to join the current herb set due to some herb compatibility rules.  Different from BPR, multi-label loss computes the distance between the recommended herb set with the ground truth herb set, which evaluates the results from the set view.  It is also the reason we do not add the positive-negative label margin constraint \cite{Zhang2006MultilabelNN} into our loss function in (\ref{loss_function}). 

\begin{table}[!tb]
\caption{Comparison of different loss functions }
\centering
\setlength{\tabcolsep}{0.8mm}{
\begin{tabular}{|c|c|c|c|c|c|c|}
\hline
Approaches                                                              & p@5    & p@20   & r@5    & r@20   & ndcg@5 & ndcg@20 \\ \hline
\begin{tabular}[c]{@{}c@{}}NGCF w/ SI\\ BPR\end{tabular}          & 0.2760 & 0.1606 & 0.1953 & 0.4472 & 0.3825 & 0.5624  \\ \hline
\begin{tabular}[c]{@{}c@{}}Bipar-GCN w/ SI\\ BPR\end{tabular}        & 0.2774 & 0.1623 & 0.1951 & 0.4479 & 0.3762 & 0.5565  \\ \hline
\begin{tabular}[c]{@{}c@{}}NGCF w/ SI\\ multi-label\end{tabular}     & 0.2787 & 0.1634 & 0.1933 & 0.4505 & 0.3790 & 0.5599  \\ \hline
 \begin{tabular}[c]{@{}c@{}}Bipar-GCN w/ SI\\ multi-label\end{tabular} & \textbf{0.2914} & \textbf{0.1690} & \textbf{0.2060} & \textbf{0.4695} & \textbf{0.3885} & \textbf{0.5699} \\ \hline
\end{tabular}}

    \label{tab:loss_function}
\end{table}

\subsubsection{Case Study} (RQ5) 
 
\begin{figure}[!tb]
  \centering
  \includegraphics[width=1.0\linewidth]{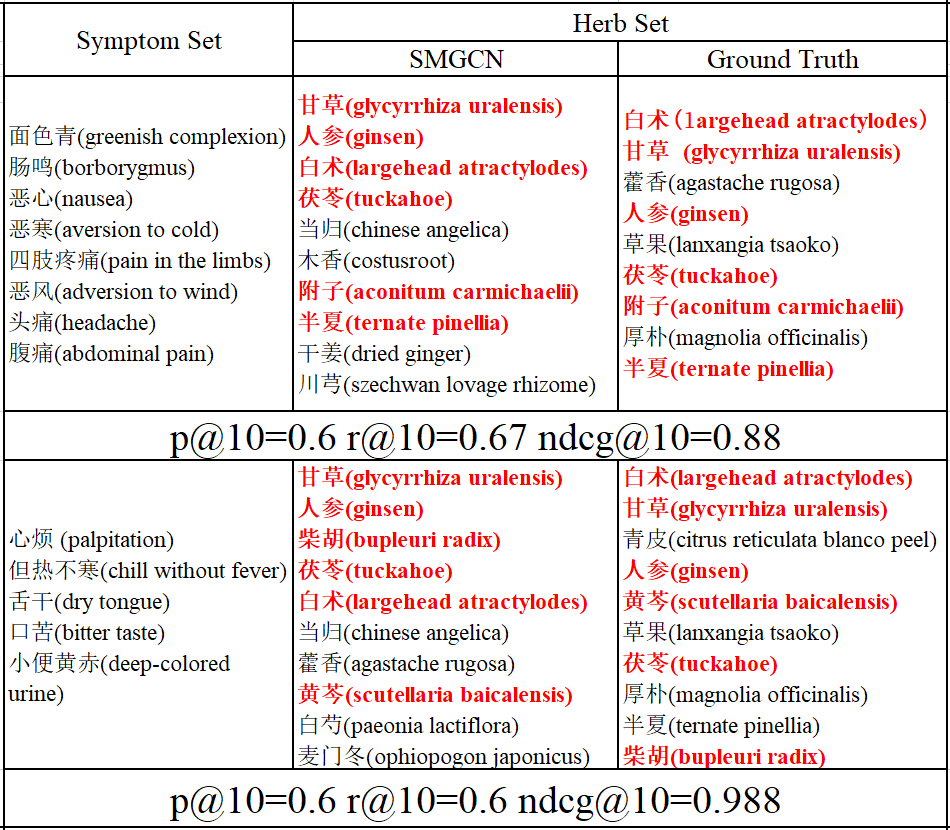}
  \caption{The herb recommendation cases.}
  \label{fig:case}
\end{figure}

In this part, we conduct a case study to verify the rationality of our proposed herb recommendation approach.  
Fig.~\ref{fig:case} shows two real examples in the herb recommendation scenario.  Given the symptom set, our proposed SMGCN generates a herb set to cure the syndrome with the listed symptoms. 
In the \textbf{Herb Set} column, the bold red font indicates the common herbs between the herb set recommended by SMGCN and the ground truth.   According to the herbal knowledge, the missing herbs actually have similar functions with the remaining ground-truth herbs and can be alternatives in clinical practice.  Through the above comparative analysis, we can find that our proposed SMGCN has the ability to provide reasonable herb recommendations. 

\section{RELATED WORK}
\subsection{Herb Recommendation}
Prescriptions play a vital role in TCM inheritance of clinical experience and practice.  The development history of TCM prescription mining contains three stages: 1) 
\emph{traditional frequency statistic} and \emph{data mining techniques}, mainly including association analysis, simple clustering, and classification methods; 2) \emph{topic models}.  Existing researches \cite{ma2016discovering,Fan2016TCM,Wang2016ACP,Chen2018HeterogeneousIN,Ruan2017THClusterHS,Yao2018ATM,ji2017latent,Wang2019AKG} compute the conditional probability of the co-occurred symptom and herb words to capture the relations among symptoms and herbs; and 3) \emph{graph model-based methods. }
Studies \cite{li2018exploration,li2017distributed,Ruan2019DiscoveringRF,ruan2019exploring} organize TCM prescriptions into graphs to capture the complex regularities.  Because the methods in the first category are only suitable for a single disease, we mainly focus on the second and third categories. 

\textbf{Topic Model Based Herb Recommendation. } 
Topic models are applied to process prescriptions in natural languages, where TCM prescriptions are documents containing herbs and symptoms as words.  The beneath motivation is that herbs and symptoms occurring under the same topic are similar. 
Ma et al. \cite{ma2016discovering} propose a  
 ``symptom-syndrome'' model to mine the correlation between symptoms and latent syndrome topics. 
Ji et al. \cite{ji2017latent} consider ``pathogenesis" as the latent topics to connect symptoms and herbs. 
 Lin et al. \cite{Fan2016TCM} jointly model symptoms, herbs, diagnosis, and treatment in prescriptions through topic models. 
 Wang et al. \cite{Wang2016ACP} design an asymmetric probability generation model to model symptoms, herbs, and diseases simultaneously. 
Yao et al. \cite{Yao2018ATM} integrate TCM concepts such as ``syndrome", ``treatment," and ``herb roles" into topic modeling, to better characterize the generative process of prescriptions.  
Chen et al. \cite{Chen2018HeterogeneousIN} and Wang et al. \cite{Wang2019AKG} introduce TCM domain knowledge into topic models to capture the herb compatibility regularities. 
 
Unfortunately, standard topic models are not very friendly to short texts.
Thus, the sparsity of prescriptions \cite{Ruan2019DiscoveringRF} will limit the performance of topic models on large-scale prescriptions to some extent. 
Besides, they cannot analyze the complex interrelationships among various entities comprehensively.

\textbf{Graph Based Herb Recommendation. } 
A graph is an effective tool to model complex relation data.  Graph representation learning-based herb recommendation is a hot research topic nowadays, which mainly focuses on obtaining the low-dimensional representations of TCM entities, and then recommends herbs based on the embeddings.  Some researches have introduced deep learning techniques into graph-based prescription mining.  Li et al. \cite{li2018exploration} utilize the attentional Seq2Seq \cite{Zhang2019Seq2SeqAS}  to design a multi-label classification method, in order to automatically generate prescriptions. 
 Li et al. \cite{li2017distributed} adopt the BRNN \cite{Schuster1997BidirectionalRN} to do text representation learning for the herb words in the TCM  literature for treatment complement task. 
 \cite{Ruan2019DiscoveringRF, ruan2019exploring} integrate the autoencoder model with meta-path to mine the  TCM heterogeneous information network.

The weak point of the above graph-based models is that the applied deep learning techniques are initially designed for the euclidean space data and lack the interpretability and reasoning ability for the non-euclidean space graph data.

\subsection{Graph Neural Networks-based recommender systems}
Graph neural networks  (GNNs)  are the extension of neural networks on the graph data, which can handle both node features and edge structures of graphs simultaneously.  
Due to its convincing performance and high interpretability, GNNs have been widely applied in recommender systems recently.  GNNs are applied to different kinds of graphs as follows:
1) \emph{User-item Interaction Graphs}:
Berg et al. \cite{berg2017graph} present a graph convolutional matrix completion model based on the auto-encoder framework. 
Wang et al. \cite{wang2019neural} encode the collaborative signal in the embedding process based on GNN, which can capture the collaborative filtering effect sufficiently;
2) \emph{Knowledge Graphs}:
Wang et al. \cite{Wang2018RippleNetPU} propose the Ripple Network,  which iteratively extends a user's potential interests along edges in a knowledge graph to stimulate the propagation of user preferences.  
 Wang et al. \cite{wang2019kgat} propose Knowledge Graph Attention Network, which recursively propagates the embeddings from a node's neighbors to obtain the node embedding, and adopts the attention mechanism to discriminate the importance of the neighbors;
3) \emph{User Social Networks}:
 Wu et al. \cite{wu2018socialgcn} and  Fan et al. \cite{fan2019graph} apply  GCNs to capture how users' preferences are influenced by the social diffusion process in social networks;  
4) \emph{User Sequential Behavior Graphs}:
Wu et al. \cite{wu2019session} and Wang et al. \cite{wang2020gf} apply GNN for  session-based recommendation by capturing complex transition relations between items in user behavior sequences.

\section{CONCLUSION AND FUTURE WORK}

In this paper, we investigate the herb recommendation task from the novel perspective of taking implicit syndrome induction into consideration.  
We develop a series of GCNs to simultaneously learn the symptom embedding and herb embedding from the symptom-herb, symptom-symptom, and herb-herb graphs. 
To learn the overall implicit syndrome embedding, we feed multiple symptom embeddings into an MLP, which is later integrated with the herb embeddings to generate herb recommendation. 
The extensive experiments carried out on a public TCM dataset demonstrate the superiority of the proposed model, validating the effectiveness of mimicking the syndrome induction by experienced doctors. 

In future work, for embedding learning, we will improve the embedding quality of the TCM entities by adopting advanced techniques such as the attention mechanism.  
For graph construction, we will introduce more TCM domain-specific knowledge, including dosage and contraindications of herbs into the TCM graphs. 



\bibliographystyle{IEEEtran}
\bibliography{IEEEabrv,ref.bib}

\end{document}